\documentclass[twocolumn]{aastex631}

\usepackage{CJK}
\usepackage{graphicx}
\usepackage{amsmath}

\received{mm dd, yyyy}
\revised{mm dd, yyyy}
\accepted{mm dd, yyyy}
\submitjournal{APJ}

\begin{document}

\begin{CJK*}{UTF8}{gbsn}

\title{Unveiling the galactic baryon cycle process by an empirical model}

\correspondingauthor{Xi Kang}
\email{kangxi@zju.edu.cn, chenyaoxin@zju.edu.cn}

\author[0000-0002-8694-0903]{Yaoxin Chen(陈瑶鑫)}
\affiliation{Institute for Astronomy, School of Physics, Zhejiang University, Hangzhou 310027, China}

\author[0000-0002-5458-4254]{Xi Kang(康熙)}
\affiliation{Institute for Astronomy, School of Physics, Zhejiang University, Hangzhou 310027, China}
\affiliation{Center for Cosmology and Computational Astrophysics, Zhejiang University, Hangzhou 310027, China}
\affiliation{Purple Mountain Observatory, 10 Yuan Hua Road, Nanjing 210034, China}

\author[0000-0003-4936-8247]{Hong Guo(郭宏)}
\affiliation{Shanghai Astronomical Observatory, Chinese Academy of Sciences, Shanghai 200030, China}

\author[0000-0001-8426-9493]{Hou-Zun Chen(陈厚尊)}
\affiliation{Institute for Astronomy, School of Physics, Zhejiang University, Hangzhou 310027, China}

\author[0000-0002-8817-4587]{Jiafeng Lu(卢家风)}
\affiliation{Institute for Astronomy, School of Physics, Zhejiang University, Hangzhou 310027, China}

\begin{abstract}
We propose an empirical model to describe and constrain the baryon cycle process during galaxy evolution. This model utilizes the evolution of star formation rate, derived from the stellar mass-halo mass relations (SHMRs) across different redshifts, and the cold gas content, derived from the NeutralUniverseMachine model, to constrain gas accretion and recycle of gas outflow in the model galaxy. Additionally, through detailed modeling of each cycle process, particularly the recycling process, and utilizing the gas-phase mass-metallicity relation (MZR) at $z=0$ as a constraint, our model establishes a relation between the recycle fraction and halo mass. It is found that the fraction of gas recycled from the outflow is a function of halo mass, with a value of $25\%$ in galaxies with halo mass $\sim10^{10.4}M_{\rm \odot}$, increasing to $75\%$ in halos with mass $\sim 10^{12}M_{\rm \odot}$. We also find that the mass loading factor from the FIRE-2 simulation matches well with the constraints from both observational data and our model. Furthermore, using the gas content and metallicity of the circumgalactic medium (CGM) obtained from hydrodynamical simulations as constraints, our model predicts that on average $70\%$ of universal baryon accretion is accreted to the halo and $80\%$ of the non-recycled gas in the outflow has escaped from the galaxy, entering the intergalactic medium (IGM). However, we note that future observational data are needed to finally constrain the mass and metal exchange between the CGM and the IGM.
\end{abstract}

\keywords{Galaxy evolution(594) --- Galaxy chemical evolution(580) --- Circumgalactic medium(1879) --- Galaxy accretion(575) --- Galactic winds (572) --- Galaxy fountains(596)}

\section{Introduction} \label{sec:intro}
Galaxies live in complex ecosystem in which baryons, including stars, gas, and metals, are exchanged between galaxies and their surrounding circumgalactic medium (CGM) and the intergalactic medium (IGM) \citep[e.g.][]{2017ARA&A..55..389T,2023ARA&A..61..131F}. Great efforts have been made to understand this complicated baryonic cycle process. Fresh gas accretion from the IGM provides raw material for star formation in the galaxy \citep[e.g.][]{2005MNRAS.363....2K,2009ApJ...694..396B,2009Natur.457..451D,2011MNRAS.417.2982F,2011MNRAS.414.2458V,2018MNRAS.478..255C}.
Stellar feedback and winds blow gas out of the galaxy, forming gas outflow and enriching metal content in the CGM \citep[e.g.][]{2015MNRAS.454.2691M,2017MNRAS.468.4170M,2018MNRAS.474.4279M,2018ApJ...867..142C,2020MNRAS.494.3971M,2021MNRAS.508.2979P}. 
Part of the outflow gas might be re-accreted into the galaxy under the influence of gravitational potential and participate in subsequent star formation, and this process is often known as 'wind recycling' \citep[e.g.][]{2010MNRAS.406.2325O,2016ApJ...824...57C,2017MNRAS.470.4698A,2019MNRAS.485.2511T,2019MNRAS.490.4786G,2020MNRAS.497.4495M}. In short, the baryon cycle process regulates the star formation and evolution of galaxies, making it a topic of significant worth for investigation.

Both theoretical and observational studies have made great progress in understanding the baryon cycle process in the last decades.
Recent studies on gas inflow have focused primarily on the theoretical aspect, particularly using hydrodynamical simulations. It is generally believed that the fraction of gas/baryon accreted by the host dark matter halo of a galaxy is lower than the universal baryon fraction, and this effect is more prominent in low-mass halos as found in semi-analytical models \citep[e.g.][]{2011MNRAS.413..101G,2015MNRAS.451.2663H,2016MNRAS.461.1760H} and hydrodynamical simulations \citep[e.g.][]{2018MNRAS.478..255C}. One possible explanation for this phenomenon is the reionization effect \citep[e.g.][]{2000ApJ...542..535G,2008MNRAS.390..920O}, that the extragalactic UV/X-ray background radiation heats the surrounding gas, thereby inhibiting gas accretion into low-mass halos with shallow potential. However, due to the complex interplay between the accreted gas and the pre-existing gas in the CGM, it remains a significant challenge in quantifying the amount of gas that can be accreted from the IGM to galaxy scale. 
In a comprehensive study, \cite{2018MNRAS.478..255C} presents the accretion rates onto galaxies derived from the EAGLE simulation and develops a physically motivated model based on the two modes of accretion: the hot and cold mode. 
However, it remains important to emphasize that gas accretion onto a galaxy is strongly affected by both stellar feedback \citep{2023MNRAS.524.4091B} and active galactic nucleus (AGN) feedback. 
Both hydrodynamical simulations \citep[e.g.][]{2015MNRAS.454.2691M,2017MNRAS.470.4698A,2018MNRAS.474.4279M,2019MNRAS.485.2511T,2019MNRAS.483.3363H,2020MNRAS.494.3971M,2021MNRAS.508.2979P,2023MNRAS.525.5868H,2023MNRAS.525.5388B} and observations \citep[e.g.][]{2017MNRAS.469.4831C,2019ApJ...885..100L,2019ApJ...886...74M,2022NatAs...6..647C,2024ApJ...966..129K} have made significant progress in studying gas outflows from galaxies, particularly focusing on the ratio between outflow mass to the star formation rate, the so-called mass loading factor. It is widely found that the mass loading factor derived from observations is generally lower than that predicted by hydrodynamical simulations \citep[e.g.][]{2019ApJ...886...74M,2022NatAs...6..647C,2024ApJ...966..129K}, although high-resolution simulations of dwarf galaxies \cite[e.g.][]{2019MNRAS.483.3363H,2023MNRAS.526.1408S,2024ApJ...960..100S} do predict a lower mass loading factor. The discrepancy in the mass loading factors across different simulations arises from distinct feedback models employed. Observational studies on gas recycling also face significant challenges. Recently, \cite{2023Sci...380..494Z,2023ApJ...952..124Z} have provided observational evidence of gas recycling. Using hydrodynamical simulations, a few studies have investigated the recycle fraction and the recycle timescale of the outflow gas \citep{2016ApJ...824...57C,2017MNRAS.470.4698A,2019MNRAS.485.2511T}. Concurrently, in semi-analytical models of galaxy formation, the treatment of gas recycling process remains at an empirical level, potentially failing to accurately capture the true nature of this phenomenon.
In summary, most key parameters that govern baryon cycle processes, particularly those characterizing gas accretion and recycling, are not directly observable.

We have developed an empirical approach in \cite{2023MNRAS.519.1899C} to investigate the evolution of galaxies and constrain the baryon cycle processes. 
We firstly determine the star formation histories (SFHs) of galaxies under the assumption that galaxies always follow the stellar mass-halo mass relations (SHMRs) across various redshifts and utilize a chemical evolution model to constrain key processes in their baryon cycle. 
Actually, the chemical evolution model in our empirical approach follows the core framework of the gas regulator model (or analytical model) \citep[e.g.][]{2010ApJ...718.1001B,2012MNRAS.421...98D,2013ApJ...772..119L,2015MNRAS.452.1184M,2010A&A...514A..73S,2017A&A...599A...6S,2018MNRAS.474.1143L,2018MNRAS.481.4000L,2020ApJ...897...81L}. However, in our work, the evolution of star and cold gas of galaxies are derived and fixed from well-established scaling relations. This enables us to focus on characterizing the essential baryon cycle processes that regulate galaxy evolution, particularly the gas recycle process.
In \cite{2023MNRAS.519.1899C}, we have found that in addition to the cold accretion due to halo growth, a portion of the outflow gas must be recycled to sustain the derived SFHs of galaxies. 
We have also made a rough estimate of the total recycle fraction $f_\mathrm{REC}$ and the recycle time $t_\mathrm{REC}$ for the outflow gas. It is found that in our fiducial model around $20-60\%$ of the outflow is recycled on a timescale of $\sim0.5-4Gyrs$, while simulations predict a slightly higher recycle fraction and a shorter recycle time \citep[e.g.][]{2015MNRAS.454.2691M,2016ApJ...824...57C,2019MNRAS.485.2511T}. 
One main caveat in \cite{2023MNRAS.519.1899C} is that the modeling of cold gas content in galaxies and their evolution with redshift is affected by the assumption of how star formation efficiency evolves with redshift, which is currently not well constrained by the observational data. In addition, the parameters $f_\mathrm{REC}$ and $t_\mathrm{REC}$ in \cite{2023MNRAS.519.1899C} are the average gas recycle fraction and recycle time for halos at $z=0$, without proper consideration of the evolution with time and halo mass.

In this work, we follow the framework of \cite{2023MNRAS.519.1899C}, but introduce two significant improvements that make the model more complete and reliable. Firstly, the cold gas content of a galaxy and its evolution is now given by an empirical model called NeutralUniverseMachine,
recently proposed by \cite{2023ApJ...955...57G}. This empirical model well reproduces the observations of both $\mathrm{HI}$
and $\mathrm{H_2}$ in the redshift range of $0<z<6$. Secondly, we now model the recycling process with a more reasonable consideration of the mass and redshift dependence for the recycling parameters.

The structure of this paper is organized as follows. In Section \ref{sec:method}, we provide a detailed description of the method for modeling galaxy evolution. The results of our model are presented in Section \ref{sec:results}, including the evolution of cold gas (Section \ref{subsec:result_gas_evolution}), the best-fit results for parameters in gas cycle processes (Section \ref{subsec:result_best_fitting_f_rec} and Section \ref{subsec:result_best_fitting_acc}) and some predictions of our model (Section \ref{subsec:result_prediction}). 
Conclusions are presented in Section \ref{sec:summary}.
Throughout this paper, we assume a standard flat $\Lambda$CDM cosmology with $\Omega_\mathrm{m}=0.315, \Omega_\Lambda=0.685$, $f_\mathrm{b}=\Omega_\mathrm{b}/\Omega_\mathrm{m}=0.15$ and $H_0=67.3 kms^{-1}Mpc^{-1}$, and
a \citet{2003PASP..115..763C} initial mass function (IMF).

\section{Method} \label{sec:method}
The details of our method are described in this section. 
Section \ref{subsec:method_overview} gives a brief overview of our method.
Then several main ingredients of our method are introduced in the following sections separately: how to derive the star formation histories (SFHs) of our model galaxies in Section \ref{subsec:method_sfh}, how to get the cold gas evolution of galaxies in Section \ref{subsec:method_cold_gas_evolution}, the chemical evolution model in Section \ref{subsec:method_chemical}, and how to model and constrain baryon cycle processes in Section \ref{subsec:method_constrain_baryon}.

\subsection{Overview}\label{subsec:method_overview} 

In this section, we briefly introduce the overall framework and modeling process of our empirical method.  We note that although all details have been presented in \cite{2023MNRAS.519.1899C}, we outline them here for easy access of the readers and the completeness of this paper. 

We select $10$ model galaxies with their dark matter halo mass ($M_{\mathrm{h}}$) evenly distributed between $10^{11}M_\odot$ and $10^{12}M_\odot$ at $z=0$. For convenience, we denote the model galaxy with a halo mass of $(i+1)\times10^{11}M_\odot$ at $z=0$ as $Gi$, where $i=0, 1, ..., 9$. 
This sample size is statistically justified as our model focuses exclusively on the mean mass-dependent galactic properties, treating all galaxies of the same mass identically regardless of their individual formation histories or environmental influences, and explicitly neglecting the scatter in these scaling relations. In addition, we focus on galaxies within this mass range for the following reason. Our model neglects halo mergers (also galaxy mergers) to minimize the uncertainties arising from modeling galaxy mergers and associated star formation. Previous studies have shown that for galaxies with $M_{\ast}<10^{11}M_{\odot}$ or $M_{\mathrm{h}}<10^{12}M_{\odot}$, galaxy mergers contribute very little to the stellar mass growth of galaxies \citep[e.g.][]{2009ApJ...696..620C, 2013ApJ...770...57B}. 
Therefore, it is quite reasonable to neglect galaxy merger and attribute the growth of stellar mass primarily to star formation in the model.

The star formation histories (SFHs) of galaxies are determined by combining the halo mass growth histories ($M_{\mathrm{h}}(z)$) and the stellar mass-halo mass relations (SHMRs) at different redshifts ($M_{\ast}(M_{\mathrm{h}},z)$, which was initially proposed by \cite{2009ApJ...696..620C}. 
In this work, the halo mass growth histories, $M_{\mathrm{h}}(z)$, are derived by integrating the mean halo mass accretion rate from the two Millennium simulations, with more details in Section \ref{subsec:method_sfh}. Taking galaxy $G0$ as an example, its progenitor halo mass at a given redshift $z$ is $M_{\mathrm{h}}(z)$, we use the SHMR at that redshift to get its stellar mass $M_{\ast}(z)$. By performing the same procedure at a slightly later time step, say $z+dz$, we can then obtain a new stellar mass $M_{\ast}(z+dz)$ based on its new halo mass, $M_\mathrm{h}(z+dz)$ at $z+dz$. By ascribing the stellar mass growth to star formation in the galaxy $G0$, it is straight to get the evolution of its star formation rate.

The evolution of cold gas in galaxies is characterized by an empirical model for the evolution of $\mathrm{HI}$ and $\mathrm{H_2}$ gas along with dark matter halos, named “NeutralUniverseMachine”, recently proposed by \cite{2023ApJ...955...57G}. 
The NeutralUniverseMachine model takes the star formation rate and halo mass of the galaxy as input to predict the $\mathrm{HI}$ and $\mathrm{H_2}$ gas mass. The details of the calculation are presented in Section \ref{subsec:method_cold_gas_evolution}.

Once the SFHs and cold gas evolution of galaxies are determined, we can investigate the gas cycle processes and metal enrichment history by utilizing a simple chemical evolution model described in Section \ref{subsec:method_chemical}. This model involves several key physical processes, including gas accretion, star formation, gas outflow, and recycle of outflow gas.
In Section \ref{subsec:method_constrain_baryon}, we delve into detailed modeling of those baryon cycle processes, particularly focusing on gas recycling and fresh gas accretion. By incorporating more robust results from simulations and observations, such as the mass loading factor and the gas-phase mass-metallicity relation (MZR), we are able to constrain the physical parameters of gas cycle, especially those associated with gas recycling and fresh gas accretion.
Note that when we constrain the parameters of baryon cycle, we do consider galaxies with lower mass (down to $M_\mathrm{h}(z=0)= 3\times 10^{10}M_\odot$) to better constrain the parameters at low-mass end. However, our main focus remains on the behaviors of galaxies within the above-mentioned mass range.

\subsection{Star formation histories (SFHs)}
\label{subsec:method_sfh}

In this work, we use the stellar mass-halo mass relation (SHMR) obtained using the abundance matching technique by \citet{2020A&A...634A.135G}. The introduction to the abundance matching technique and the progress of SHMR is beyond the scope of this paper, and the readers are referred to the review paper by \citet{2018ARA&A..56..435W}.
\citet{2020A&A...634A.135G} determined the SHMRs up to $z=4$ using the COSMOS data, and they adopted the simple double power-law function proposed by \citet{2010ApJ...710..903M},
\begin{equation}
    \frac{M_*}{M_\mathrm{h}}(z)=2A(z)\left[\left(\frac{M_\mathrm{h}}{M_\mathrm{A}(z)}\right)^{-\beta(z)}+\left(\frac{M_\mathrm{h}}{M_\mathrm{A}(z)}\right)^{\gamma(z)}\right]^{-1},
	\label{eq:SHMR}
\end{equation}
where $A$ is the normalization at the characteristic halo mass $M_\mathrm{A}$, while $\beta$ and $\gamma$ are the slopes of the relation at low- and high- mass ends respectively. We use the best-fit parameters in Table 4 of \citet{2020A&A...634A.135G} and linearly extrapolate their relations to a lower halo/galaxy mass if needed. The scatter in stellar mass at a given halo mass is a function of halo mass \citep[e.g.][]{2010ApJ...717..379B}, and in our work we follow \citet{2020A&A...634A.135G} to use a typical constant with $\sigma_R = 0.2$ dex to their best-fit SHMRs.

The halo mass growth histories are derived by integrating the mean halo mass accretion rate in \cite{2010MNRAS.406.2267F}. 
\cite{2010MNRAS.406.2267F} quantifies the mass growth rates of dark matter halos by analyzing data from the Millennium and Millennium-II simulations. The high resolution of the Millennium-II simulation allows for halo formation history to very low-mass dark matter halos ($\sim 10^{10}M_\odot$), effectively covering the range of halo mass that is of interest to us. We adopt the following mean halo mass accretion rate,
\begin{equation}
    \begin{aligned}
    \langle \frac{dM_\mathrm{h}}{dt} \rangle(z) =\ &46.1M_\odot yr^{-1}\left(\frac{M_{\mathrm{h}}}{10^{12}M_\odot}\right)^{1.1}\\&\times(1+1.11z)\sqrt{\Omega_{\mathrm{m}}(1+z)^3+\Omega_\Lambda}.
	\label{eq:halo_growth}
    \end{aligned}
\end{equation}
For each model galaxy, from $G0$ to $G9$, we integrate the mean halo mass accretion rate in Equation \ref{eq:halo_growth} to derive the halo mass growth history. Although the method used to derive halo mass growth histories in this study differs from \cite{2023MNRAS.519.1899C}, there are no significant differences in the obtained results. Therefore we choose not to include a separate figure here. By combining the halo mass growth histories with the SHMRs at different redshifts (Equation \ref{eq:SHMR}), the stellar mass growth histories $\dot{M}_*(t)$ can be obtained by following the steps outlined in Section \ref{subsec:method_overview}. Since the SFHs obtained in this work are almost the same as in Figure 2 in \cite{2023MNRAS.519.1899C}, no further analysis or additional comparison will be presented here. In summary, the findings on the evolution of the halo/stellar mass and SFHs align broadly with previous relevant studies \citep[e.g.][]{2009ApJ...696..620C,2013MNRAS.428.3121M,2013ApJ...770...57B}.

Now we have obtained the change rate of stellar mass $\dot{M}_*$ based on SHMR and halo mass growth histories. To account for mass loss from stellar evolution processes, we relate the derived stellar mass growth to the instantaneous star formation rate (SFR) through:
\begin{equation}
    \dot{M}_*(t)=(1-R)\cdot \mathrm{SFR}(t),
	\label{eq:SF}
\end{equation}
where $R$ is the returned mass fraction from evolving stellar population. $R$ is mainly determined by the adopted initial mass function (IMF) and can be calculated using the method given by \citet{2016MNRAS.455.4183V}. For our selected Chabrier IMF \citep{2003PASP..115..763C}, we take $R=0.44$.

\subsection{NeutralUniverseMachine model} \label{subsec:method_cold_gas_evolution}

In our work, the cold gas content and evolution in each model galaxy is determined by the empirical model, NeutralUniverseMachine \citep{2023ApJ...955...57G}. 
This empirical model is capable of reproducing the observations of both $\mathrm{HI}$ and $\mathrm{H_2}$ in the redshift range of $0<z<6$. 

In the NeutralUniverseMachine model, the gas content of $\mathrm{HI}$ and $\mathrm{H_2}$ in each galaxy mainly depends on the following quantities, $M_{\mathrm{h}}$, $M_\ast$, $\mathrm{SFR}$, $z$, and $z_{\mathrm{form}}$. The $\mathrm{HI}$ mass is described with the following equation, 
\begin{equation}
    \begin{aligned}
    &M_{\mathrm{HI}}=\frac{\kappa M_{\mathrm{h}}}{\mu^{-\alpha}+\mu^\beta}\left(\frac{1+z}{1+z_{\mathrm{form}}}\right)^\gamma\left(\frac{\mathrm{SFR}}{\mathrm{SFR_{MS,obs}}}\right)^\lambda,\\&\mu=M_{\mathrm{h}}/M_{\mathrm{crit}},\\&\mathrm{log}\kappa=\kappa_0+\kappa_1 z+\kappa_2z^2,\\&\mathrm{log}M_{\mathrm{crit}}=M_0+M_1 z+M_2 z^2,
    \end{aligned}
    \label{eq:HI}
\end{equation}
Here $z_{\mathrm{form}}$ is the halo formation time which is usually defined as the redshfit when the most massive progenitor reaches half of the final halo mass. In this equation, $\mathrm{SFR_{MS,obs}}$ is the SFR of a galaxy on the main sequence with given stellar mass, for which the star-forming main sequence described in \cite{2023MNRAS.519.1526P} is used. Equation \ref{eq:HI} contains ten free parameters: $\kappa_0,\ \kappa_1,\ \kappa_2,\ M_0,\ M_2,\ M_3,\ \alpha,\ \beta,\ \gamma,$ and $\lambda$.
The $\mathrm{H_2}$ mass is a function of $M_\ast$, $\mathrm{SFR}$, and $z$,
\begin{equation}
    \begin{aligned}
    &M_{\mathrm{H_2}}=\zeta M_\ast^\nu\left(\frac{\mathrm{SFR}}{\mathrm{SFR_{MS,obs}}}\right)^\eta,\\&\mathrm{log}\zeta=\zeta_0+\zeta_1 \mathrm{ln}(1+z)+\zeta_2 [\mathrm{ln}(1+z)]^2,
    \end{aligned}
    \label{eq:H2}
\end{equation}
Equation \ref{eq:H2} incorporates five free parameters: $\zeta_0,\ \zeta_1,\ \zeta_2,\ \mu,$ and $\eta$.
Therefore, the NeutralUniverseMachine model is characterized by 15 free parameters. To fully constrain these parameters, the following observational datasets were utilized: At $z\sim 0$, $\mathrm{HI}$ mass function from \cite{2023ApJ...955...57G}, $\mathrm{HI}$-halo mass relation from \cite{2020ApJ...894...92G}, $\mathrm{HI}$-stellar mass relation from \cite{2021ApJ...918...53G}, $\mathrm{H_2}$ mass function from \cite{2021MNRAS.501..411F}, $\mathrm{H_2}$-stellar mass relation from \cite{2017ApJS..233...22S} and $\mathrm{H_2-HI}$ mass ratio from \cite{2018MNRAS.476..875C}; $\mathrm{HI}$-stellar mass relation at $z\sim 1.1$ from \cite{2022ApJ...931L..34C}; Cosmic $\mathrm{HI}$ density at $0<z<5$ and cosmic $\mathrm{H_2}$ density at $0<z<6$ from \cite{2020ApJ...902..111W}. Utilizing the Bayesian inference algorithm MULTINEST \citep{2009MNRAS.398.1601F}, \cite{2023ApJ...955...57G} systematically explored the parameter space and presented the best-fitting model parameter in their Equations 17-22.

In Section \ref{subsec:method_sfh}, we have obtained the halo mass growth history and SFHs of each model galaxy. 
Specifically, we have acquired the following quantities for each model galaxy at any redshift: $M_{\mathrm{h}}$, $M_\ast$, $\mathrm{SFR}$, $z$, and $z_{\mathrm{form}}$.
With the above properties of galaxies at hand, we can directly use the above equations to get the $\mathrm{HI}$ and $\mathrm{H_2}$ mass for each model galaxy.

Finally, to obtain the total mass of cold gas, we have to include the contribution of other elements, in particular the dominant contribution by the helium element \footnote{In \cite{2023ApJ...955...57G}, $M_\mathrm{HI}$ represents HI mass only and $M_\mathrm{H_2}$ indicates the total mass of the molecular gas}. 
Following \cite{2023ApJ...955...57G}, the total mass of the cold gas $M_{\mathrm{gas}}$ is approximately $M_{\mathrm{gas}}\simeq 1.36M_{\mathrm{HI}}+M_{\mathrm{H_2}}$.
In Section \ref{subsec:result_gas_evolution} we will present results of the cold gas content and its evolution over time for our ten model galaxies.

\subsection{Chemical evolution model}
\label{subsec:method_chemical}
In this section, we introduce a simple chemical evolution model to describe the baryon cycle, including the gas cycle and the metallicity enrichment.

The cold gas reservoir of a galaxy can be described as, 
\begin{equation}
    \dot{M}_{\mathrm{gas}}(t)=\dot{M}_{\mathrm{gas,in}}(t)-\dot{M}_{\mathrm{gas,out}}(t)-\dot{M}_{*}(t).
	\label{eq:total}
\end{equation}
In this equation, $\dot{M}_{\ast}(t)$ represents the cold gas consumption by star formation, $\dot{M}_{\mathrm{gas,out}}(t)$ is the cold gas blown away by stellar winds and feedback, and $\dot{M}_{\mathrm{gas,in}}(t)$ is the total rate of cold gas accretion, which includes fresh gas accretion from the IGM and recycle of outflow gas ($\dot{M}_{\mathrm{gas,in}}(t)=\dot{M}_{\mathrm{gas,acc}}(t)+\dot{M}_{\mathrm{gas,rec}}(t)$). With the purpose of constraining how much gas accretion is from recycle of outflow, Equation \ref{eq:total} can be rewritten as:
\begin{equation}
    \begin{aligned}
    \dot{M}_{\mathrm{gas}}(t)&=\dot{M}_{\mathrm{gas,acc}}(t)+\dot{M}_{\mathrm{gas,rec}}(t)\\&-\dot{M}_{\mathrm{gas,out}}(t)-\dot{M}_{*}(t).
	\end{aligned}
    \label{eq:total_acc_rec}
\end{equation}

The chemical enrichment of the interstellar medium (ISM) is intrinsically linked to stellar evolution processes. As stars form and evolve, they synthesize heavy elements (metals) that are subsequently returned to the ISM through stellar feedback or stellar wind. The metal content is assumed to mix uniformly with the ISM. A yield parameter $y_{\mathrm{Z}}$ is often used to characterize this process, which is the ratio of the total mass of metals that a stellar population releases into the ISM and the remained stellar mass of that stellar population:
\begin{equation}
    y_{\mathrm{Z}}=\frac{\dot{M}_\mathrm{new\_metal}}{(1-R)\mathrm{SFR}}=\frac{\dot{M}_\mathrm{new\_metal}}{\dot{M}_*},
	\label{eq:yield}
\end{equation}
where $\dot{M}_\mathrm{new\_metal}$ represents the metal mass growth rate in ISM due to the metal production. Obviously, the value of $y_{\mathrm{Z}}$ depends on the selected IMF. For Chabrier IMF \citep{2003PASP..115..763C}, we take $y_{\mathrm{Z}}$ as 0.06 from the work by \citet{2016MNRAS.455.4183V}. For more definitions or calculation details of $R$ and $y_{\mathrm{Z}}$, we refer readers to \cite{2016MNRAS.455.4183V}. 
Similar to the evolution of cold gas reservoir (Equation \ref{eq:total_acc_rec}), the evolution of metal in the ISM can be described by the following equation:

\begin{equation}
    \begin{aligned}
    \dot{M}_\mathrm{Z,gas}(t)&=Z_{\mathrm{gas,acc}}(t)\dot{M}_{\mathrm{gas,acc}}(t)\\&+Z_{\mathrm{gas,rec}}(t)\dot{M}_{\mathrm{gas,rec}}(t)\\&-Z_\mathrm{gas,out}(t)\dot{M}_{\mathrm{gas,out}}(t)\\&+(-Z_\mathrm{gas}(t)+y_\mathrm{Z})\dot{M}_{*}(t).
    \label{eq:metal}
    \end{aligned}
\end{equation}
Here, $Z_\mathrm{gas}$ represents the metallicity of the ISM, defined as the ratio of the metal mass to the gas mass in the ISM, expressed as $Z_\mathrm{gas}=M_{\mathrm{Z,gas}}/M_{\mathrm{gas}}$.
Similarly, $Z_{\mathrm{gas,acc}}$, $Z_{\mathrm{gas,out}}$ and $Z_{\mathrm{gas,rec}}$ correspond to the metallicity in fresh accretion gas, outflow gas, and recycled gas, respectively. Moreover, it is generally believed that the metallicity of gas from fresh accretion from the IGM is primordial with $Z_{\mathrm{gas,acc}}\sim 0$. 
Then Equation \ref{eq:metal} becomes:
\begin{equation}
    \begin{aligned}
    \dot{M}_\mathrm{Z,gas}(t)&=Z_{\mathrm{gas,rec}}(t)\dot{M}_{\mathrm{gas,rec}}(t)\\&-Z_\mathrm{gas,out}(t)\dot{M}_{\mathrm{gas,out}}(t)\\&+(-Z_\mathrm{gas}(t)+y_\mathrm{Z})\dot{M}_{*}(t).
    \label{eq:metal_final}
    \end{aligned}
\end{equation}
Here, Equation \ref{eq:metal_final} indicates that the metal content in the ISM is independent of the fresh gas accretion from the IGM. 
Thus we can use the observed mass-metallicity relation of galaxies to constrain the recycle process of gas from outflow. Together with the evolution of cold gas content described by Equation \ref{eq:total_acc_rec}, we are able to constrain the parameter of fresh cold gas accretion from the IGM.
It is crucial to emphasize that our parameter constraints critically rely on assumption of primordial (zero-metallicity) IGM accretion. Allowing metal-enriched accretion would introduce strong degeneracies between accretion and recycle parameters, limiting our ability to reliably determine these parameters in the current framework. Therefore, although galactic feedback processes may have weakly polluted the IGM, rendering its gas composition no longer purely primordial (especially at low redshift), we maintain the primordial accretion assumption as other similar studies \citep[e.g.][]{2012MNRAS.421...98D,2015MNRAS.452.1184M,2017A&A...599A...6S}.

\subsection{Modeling of gas cycle and parameter fitting}\label{subsec:method_constrain_baryon}

In this section, we model the gas cycling processes described by Equation \ref{eq:total_acc_rec}, along with the evolution of metals outlined by Equation \ref{eq:metal_final}. Subsequently, we utilize observations and simulations to constrain the parameters involved in these gas cycle processes.

\begin{figure}
    \centering
    \includegraphics[width=\columnwidth]{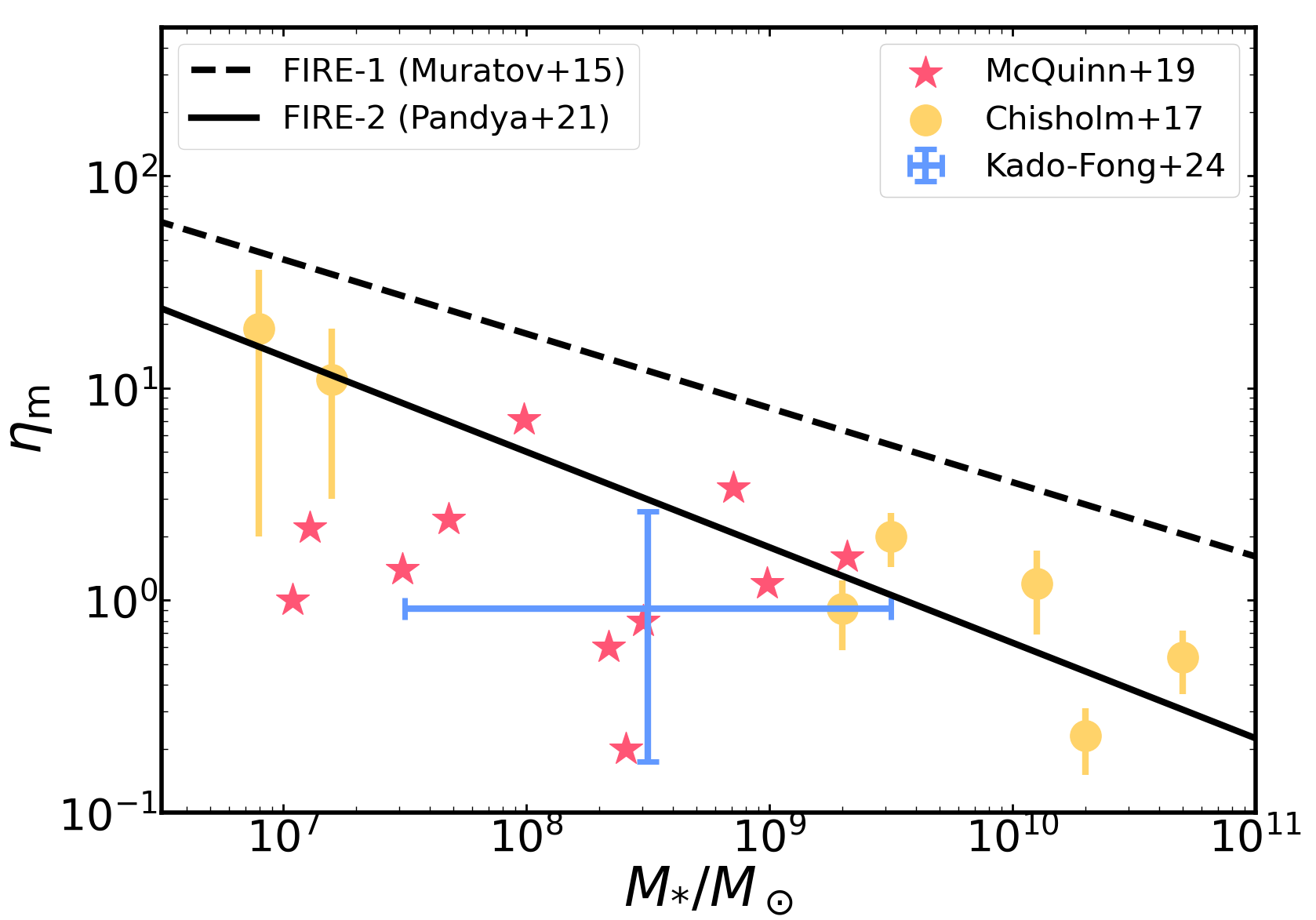}
    \caption{The input mass loading factor $\eta_\mathrm{m}(M_*)$ adopted from simulations (black lines). The dashed black and solid black lines are the fitting relations from the FIRE-1 simulation \citep{2015MNRAS.454.2691M} and FIRE-2 simulation \citep{2021MNRAS.508.2979P}, respectively. All scattered dots represent the data from observations, including \citet{2017MNRAS.469.4831C}, \citet{2019ApJ...886...74M}, \citet{2024ApJ...966..129K}. It is seen that result from the FIRE-2 simulation agrees better with the data.}
    \label{fig:Mass_loading_factor}
\end{figure}

Our modeling of the outflow gas follows a similar approach as detailed in \cite{2023MNRAS.519.1899C}. The galactic gas outflows are primarily caused by stellar feedback and winds, and the rate of gas outflow due to stellar feedback is often assumed to be proportional to the SFR with a factor depending on galaxy halo/stellar mass (or halo velocity) and redshift as,
\begin{equation}
    \dot{M}_\mathrm{gas,out}(t)=\eta_\mathrm{m}\times SFR(t)
	\label{eq:out}
\end{equation}
Here $\eta_\mathrm{m}$ is usually referred as the mass loading factor.  It is widely recognized that both the mass loading factor and the eventual fate of the outflow gas are significantly influenced by the models of stellar feedback in simulations \citep[e.g.][]{2015MNRAS.454.2691M, 2016ApJ...824...57C,2017MNRAS.472..949B,2017MNRAS.470.4698A, 2019MNRAS.485.2511T,2019MNRAS.490.3234N, 2020MNRAS.494.3971M,2021MNRAS.508.2979P}. We refer readers to the paper by \citet{2020MNRAS.494.3971M} for comparison of $\eta_\mathrm{m}$ between different cosmological simulations and semi-analytical models.
In this work, we use the mass loading factor derived from the FIRE simulation as a base to explore how our model fitting depends on the feedback parameter. However, it is found that $\eta_\mathrm{m}$ is not trivial to quantify even for a given set of FIRE simulations \citep{2021MNRAS.508.2979P}. In Figure \ref{fig:Mass_loading_factor} we show the comparison of the mass loading factors between the two FIRE simulations \citep{2015MNRAS.454.2691M,2021MNRAS.508.2979P} and a few observational data \citep{2017MNRAS.469.4831C, 2019ApJ...886...74M, 2024ApJ...966..129K}. It is seen that $\eta_\mathrm{m}$ from the FIRE-1 simulation \citep{2015MNRAS.454.2691M} is systematically higher than the observational data, while $\eta_\mathrm{m}$ from the FIRE-2 simulation \citep{2021MNRAS.508.2979P} agrees better with the data. In fact, $\eta_\mathrm{m}$ of FIRE-2 is roughly a factor of 0.3 of the FIRE-1 result. 
As pointed by \cite{2021MNRAS.508.2979P}, the difference from the FIRE-1 and FIRE-2 simulations primarily arises from the criteria used to identify outflow gas. 
In our following analysis, we adopt the mass loading factor $\eta_\mathrm{m}$ from the FIRE-2 simulation (in closer agreement with observations) as our fiducial model, and use the one from FIRE-1 as a comparative test case to explore the impacts on results. 
Additionally, we compare our fiducial mass loading factor with the results derived from \cite{2015MNRAS.452.1184M}, which also employs the gas regulator model. We find that our fiducial values agree well with the local ($z=0$) results from \cite{2015MNRAS.452.1184M} at the low-mass end ($M_*<10^9M_\odot$), but systematically exceed their reported values for more massive galaxies.

\begin{figure}
    \centering
    \includegraphics[width=\columnwidth]{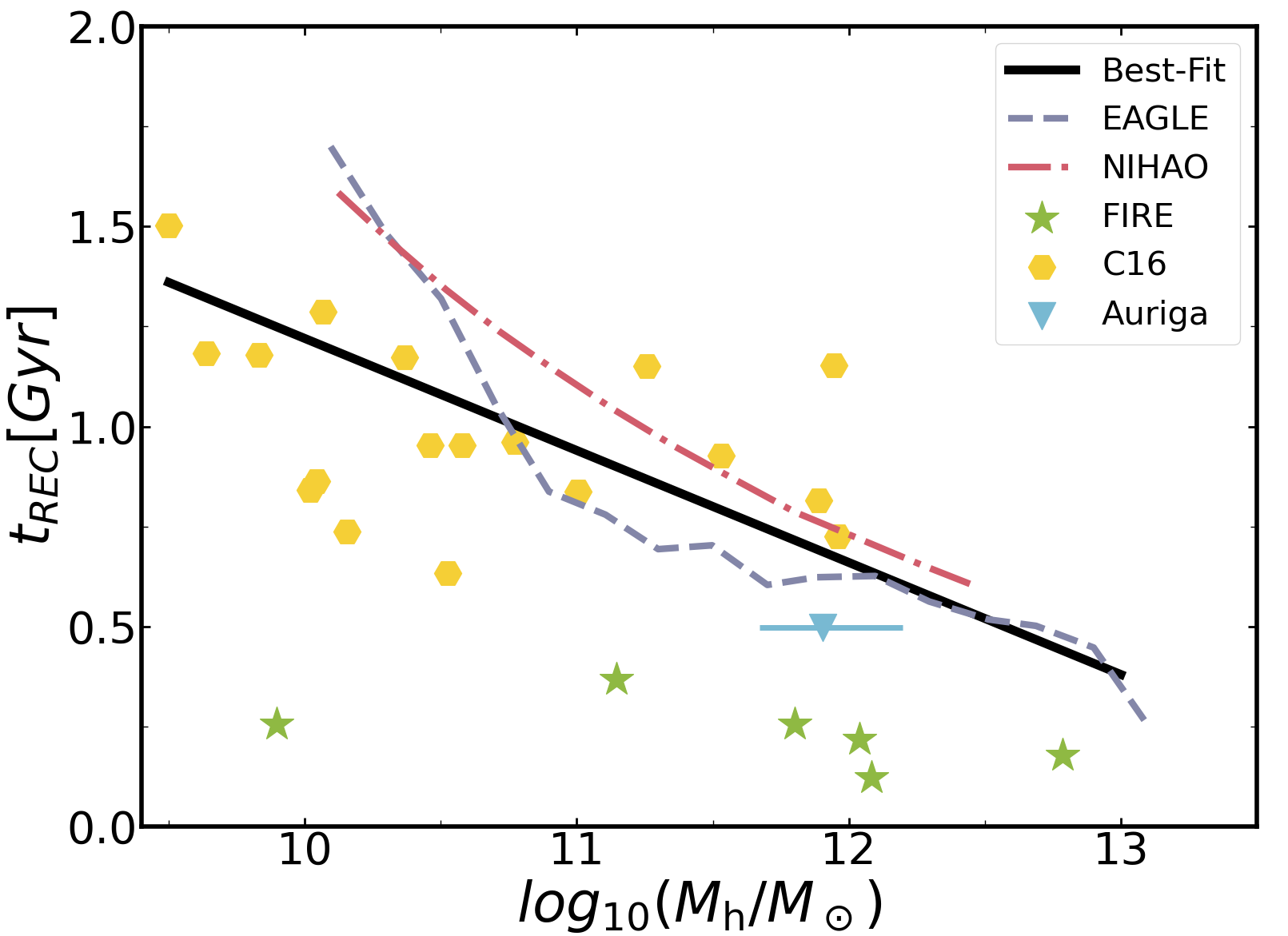}
    \caption{The input recycle time $t_\mathrm{REC}(M_\mathrm{h})$ adopted from simulations (black line). The black line represents the fitting result across multiple hydrodynamic simulations, including NIHAO simulation \citep{2019MNRAS.485.2511T}, EAGLE simulation \citep{2020MNRAS.497.4495M}, FIRE simulation \citep{2017MNRAS.470.4698A}, C16 simulation \citep{2016ApJ...824...57C} and Auriga simulation \citep{2019MNRAS.490.4786G}.}
    \label{fig:Recycle_time}
\end{figure}

As mentioned before, this work focuses on the model constraints of the gas recycling process, which are mainly characterized by two parameters, namely the recycle fraction $f_\mathrm{REC}$ and the recycle time $t_\mathrm{REC}$. However, our current definitions of these two parameters are slightly different from the ones presented in previous work in \cite{2023MNRAS.519.1899C}.
The current description of these two parameters is outlined as follows: during a time step $\Delta t$, a galaxy with halo mass $M_\mathrm{h}$ expels a specific amount of gas $\Delta M_\mathrm{gas,out}$, and a fraction $f_\mathrm{REC}$ of it, $f_\mathrm{REC} \cdot \Delta M_\mathrm{gas,out}$, will uniformly return to the galaxy over the following time interval of $t_\mathrm{REC}$.
The recycling of metals in the outflow completely follows the recycling of gas. 
To simplify our model, we assume that both the recycle fraction and the recycle time are only functions of halo mass, that is $f_\mathrm{REC}(M_\mathrm{h})$ and $t_\mathrm{REC}(M_\mathrm{h})$. 
This simplification is reasonable, considering that the recycling process of gas is mainly affected by the gravitational potential. However, we acknowledge that these two parameters may indeed depend on redshift, but that consideration is beyond the scope of our work.
Hydrodynamical simulations have also been used to study the gas recycle time and recycle fraction for the outflow. In Figure \ref{fig:Recycle_time} we show the results from a few studies. It is found that a general trend is that the recycle time is longer in low-mass halos. We choose to fit these simulation results  using a simple power law following as:
\begin{equation}
    t_{\mathrm{REC}}=\alpha_0+\alpha_1 \times \mathrm{log}_{10}M_{\mathrm{h}}.
	\label{eq:tec}
\end{equation}
And the best-fit parameters are $\alpha_0 = 4.02$ and $\alpha_1=-0.28$, shown as the black line in Figure \ref{fig:Recycle_time}.

To model the fresh gas accretion from the IGM to the galaxy, we introduce a parameter, the accretion fraction $\epsilon$, which also depends on halo mass as:
\begin{equation}
    \dot{M}_\mathrm{gas,acc}(t)=\epsilon(M_\mathrm{h})\times f_\mathrm{b} \dot{M}_\mathrm{h}(t),
	\label{eq:acc}
\end{equation}
where $f_\mathrm{b}=\Omega_\mathrm{b}/\Omega_\mathrm{m}$ is the universal baryon fraction. Here, the accretion fraction $\epsilon(M_\mathrm{h})$ in Equation \ref{eq:acc} is a combination of two effects. First, it reflects the effect that gas accreted into halo is lower than the universal baryon fraction $f_\mathrm{b}$. This is mainly due to the cosmic re-ionization effect in low-mass halos \citep{2000ApJ...542..535G} and the preventive effect near the virial radius \citep{2024MNRAS.532.3417W}. We denote the ratio of the actual gas fraction accreted into the halo to the universal baryon fraction as $\epsilon_\mathrm{halo}$, which will be further discussed in Section \ref{subsubsec:result_prediction_mass_metallicity_CGM}. The second effect, that gas entering the halo may not fully reach the central galaxy due to various feedback-related preventive effects at the CGM scale \citep[e.g.][]{2018MNRAS.473..538C,2018MNRAS.478..255C,2024MNRAS.532.3417W}, and we denote this effect as $\epsilon_\mathrm{galaxy}$. It is obvious that the accretion fraction in our equation is the product of these two effects, that is, $\epsilon=\epsilon_\mathrm{halo}\times \epsilon_\mathrm{galaxy}$.

In summary, we have employed several key parameters to characterize both galactic gas cycling and associated metal enrichment processes:
\begin{itemize}
    \item Outflow: \\Mass loading factor $\eta_\mathrm{m}$ (Input, Figure~\ref{fig:Mass_loading_factor})
    \item Recycle: \\Recycle time $t_\mathrm{REC}$ (Input, Figure~\ref{fig:Recycle_time})\\ Recycle fraction $f_\mathrm{REC}$ (Free)
    \item Fresh accretion: \\Accretion fraction $\epsilon$ (Free)
\end{itemize}
As mentioned before, mass loading factor $\eta_\mathrm{m}$ and recycle time $t_\mathrm{REC}$ are adopted from simulations, while recycle fraction $f_\mathrm{REC}$ and accretion fraction $\epsilon$ remain as free parameters requiring observational constraints.
As analyzed in Section \ref{subsec:method_chemical}, we can use the metal evolution described in Equation \ref{eq:metal_final} and the gas-phase MZR at $z=0$ to restrict the dependence of $f_\mathrm{REC}$ on $M_\mathrm{h}$ and then use Equation \ref{eq:total_acc_rec} to explore the dependence of $\epsilon$ on $M_\mathrm{h}$. To get the best-fit model parameters, we employ the Markov Chain Monte Carlo (MCMC) method to perform the non-parametric tuning of $f_\mathrm{REC}$ and $\epsilon$. 
Once again, it is worth noting that as we attempt to constrain the parameters in the baryon cycle processes here, we have indeed expanded the mass range of our model galaxies to include lower masses to better constrain the parameters at the low-mass end. However, finally our primary focus remains on the results and behaviors of parameters within the previously mentioned halo mass range ($G0$ to $G9$), which spans approximately $10^{10.5}-10^{12.0}M_\odot$. This mass range includes the halo mass of ten model galaxies from $z=2$ to $z=0$. Specifically, the halo mass of our smallest galaxy $G0$ at $z=2$ is $\sim 10^{10.5}M_\odot$, while the halo mass of our largest galaxy $G9$ at $z=0$ is $10^{12.0}M_\odot$.

\section{Results} \label{sec:results}
The main results of our work are presented in this section. 
In Section \ref{subsec:result_gas_evolution}, we show the evolution of cold gas in the ten model galaxies based on the NeutralUniverseMachine model.
The best-fit parameters for the baryon recycle fraction $f_\mathrm{REC}$ and the accretion fraction $\epsilon$ are presented in Section \ref{subsec:result_best_fitting_f_rec} and Section \ref{subsec:result_best_fitting_acc}. Based on these parameters, we also make some predictions on the metal content and its evolution in Section \ref{subsec:result_prediction}.

\subsection{The evolution of cold gas} \label{subsec:result_gas_evolution}

\begin{figure}
    \centering
    \includegraphics[width=\columnwidth]{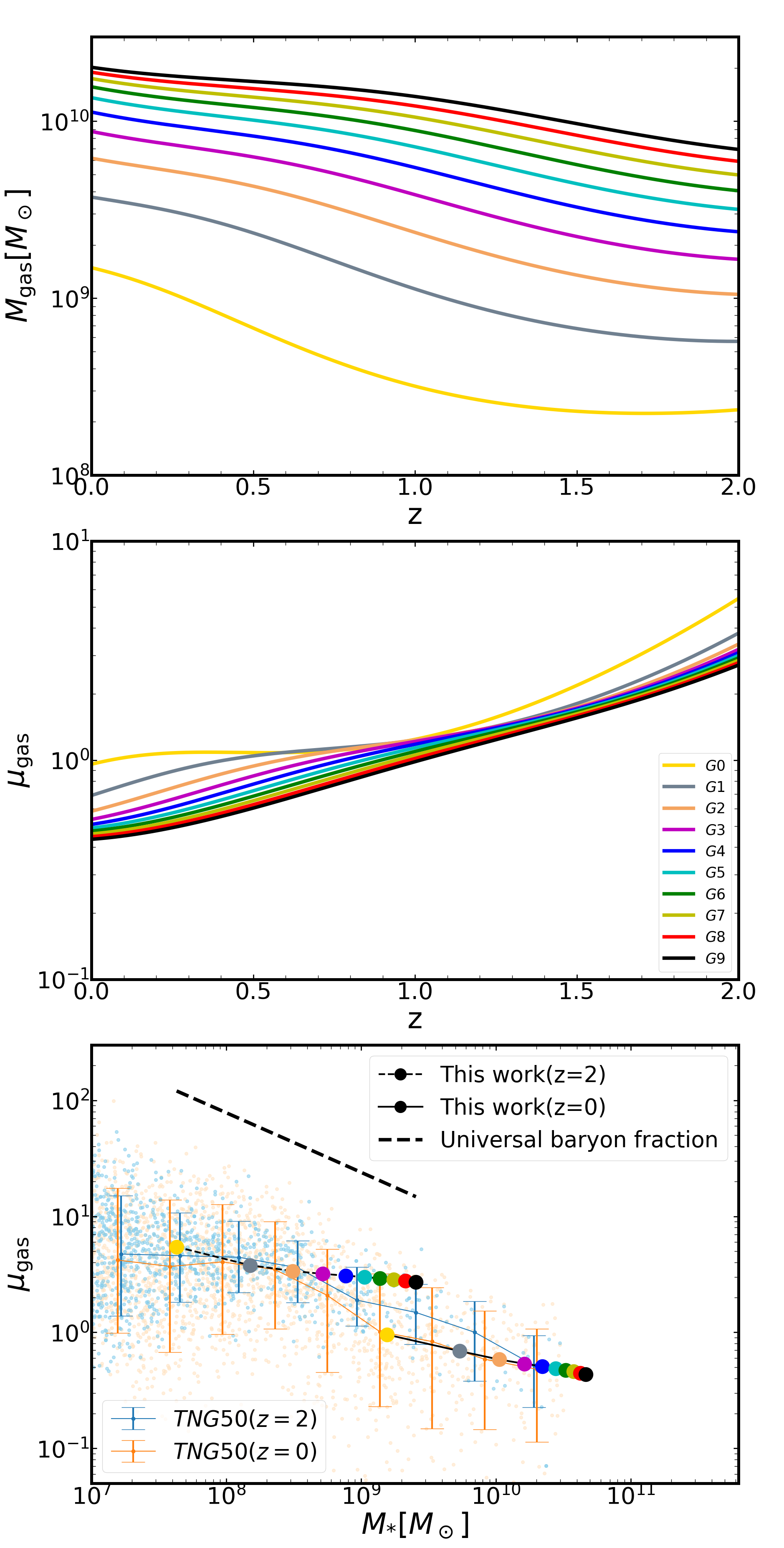}
    \caption{The prediction of cold gas for ten model galaxies obtained by NeutralUniverseMachine model \citep{2023ApJ...955...57G}. 
    The evolutions of both gas mass and gas fraction are presented in the top and middle panel, respectively. 
    Bottom panel: the gas fraction as a function of stellar mass at $z=2$ and $z=0$ (the $\mu_\mathrm{gas}-M_*$ relation). The model predictions are shown as filled color circles connected by the solid black line ($z=0$), and the dashed black line ($z=2$). The small scattered dots are from the TNG50 simulation at $z=2$ (blue dots) and at $z=0$ (orange dots), with two lines and error-bars showing the medium and the scatter.
    The thick dashed line is the upper limit of gas fraction by assuming that total baryon mass in a model galaxy is the universal baryon fraction $f_\mathrm{b}$.
    }
    \label{fig:Gas_evolution}
\end{figure}

Using the NeutralUniverseMachine model, we obtain the cold gas content and evolution for our ten model galaxies. The results are shown in Figure \ref{fig:Gas_evolution}, in which the top and middle panels show the evolution of the total cold gas, the gas fraction in the model galaxies, and the lower panel compares our model predictions with the simulation results from the TNG50. The upper panel shows that the mass of the cold gas content in all galaxies increases with time from $z=2$ to $z=0$, with a stronger redshift evolution found for low-mass galaxies.
The results indicate that in the mass range we consider in this paper, $10^{11}M_{\rm \odot}$ and $10^{12}M_{\rm \odot}$, the consumption of cold gas by star formation and feedback in the galaxy is lower than the fresh gas accretion from the IGM and the recycle of outflow gas into the galaxy.

The gas fraction, expressed as $\mu_\mathrm{gas}=M_\mathrm{gas}/M_\mathrm{*}$, is presented in the middle panel of Figure \ref{fig:Gas_evolution}. This panel reveals a continuous decline of gas fraction with time. As both stellar mass and cold gas increase with time in these model galaxies, but the gas fraction is decreasing, it indicates that our model galaxies are in delicate balance between the total gas accretion and consumption, so that the increase in stellar mass slightly surpasses the increase in cold gas.

In the bottom panel of Figure \ref{fig:Gas_evolution}, the gas fraction is plotted as a function of stellar mass at $z=0$ and $z=2$. The obtained $\mu_\mathrm{gas}-M_*$ relations are also compared with the TNG50 simulation \citep{2019MNRAS.490.3234N,2019ComAC...6....2N,2019MNRAS.490.3196P}. 
Our model galaxies at $z=0$ and $z=2$ are shown as filled color circles, connected by black solid and dashed lines, respectively. 
The small scattered dots are results of the TNG50-1 simulation \citep{2018ApJS..238...33D,2019MNRAS.487.1529D}, and the colored lines with error bars show the median and scatter (40 percentage from the median). 
Surprisingly, we find that the $\mu_\mathrm{gas}-M_*$ relations obtained for our ten model galaxies using the NeutralUniverseMachine model are in good agreement with those derived from the TNG50 simulation, both at $z=0$ and $z=2$.
The improvement over the previous result in \cite{2023MNRAS.519.1899C} arises primarily from the NeutralUniverseMachine model, which is designed to fit the observation of cold gas at various redshifts.

\subsection{Recycle fraction \texorpdfstring{$f_{\mathrm{REC}}$}{fREC}} \label{subsec:result_best_fitting_f_rec}

\begin{figure}
    \centering
    \includegraphics[width=\columnwidth]{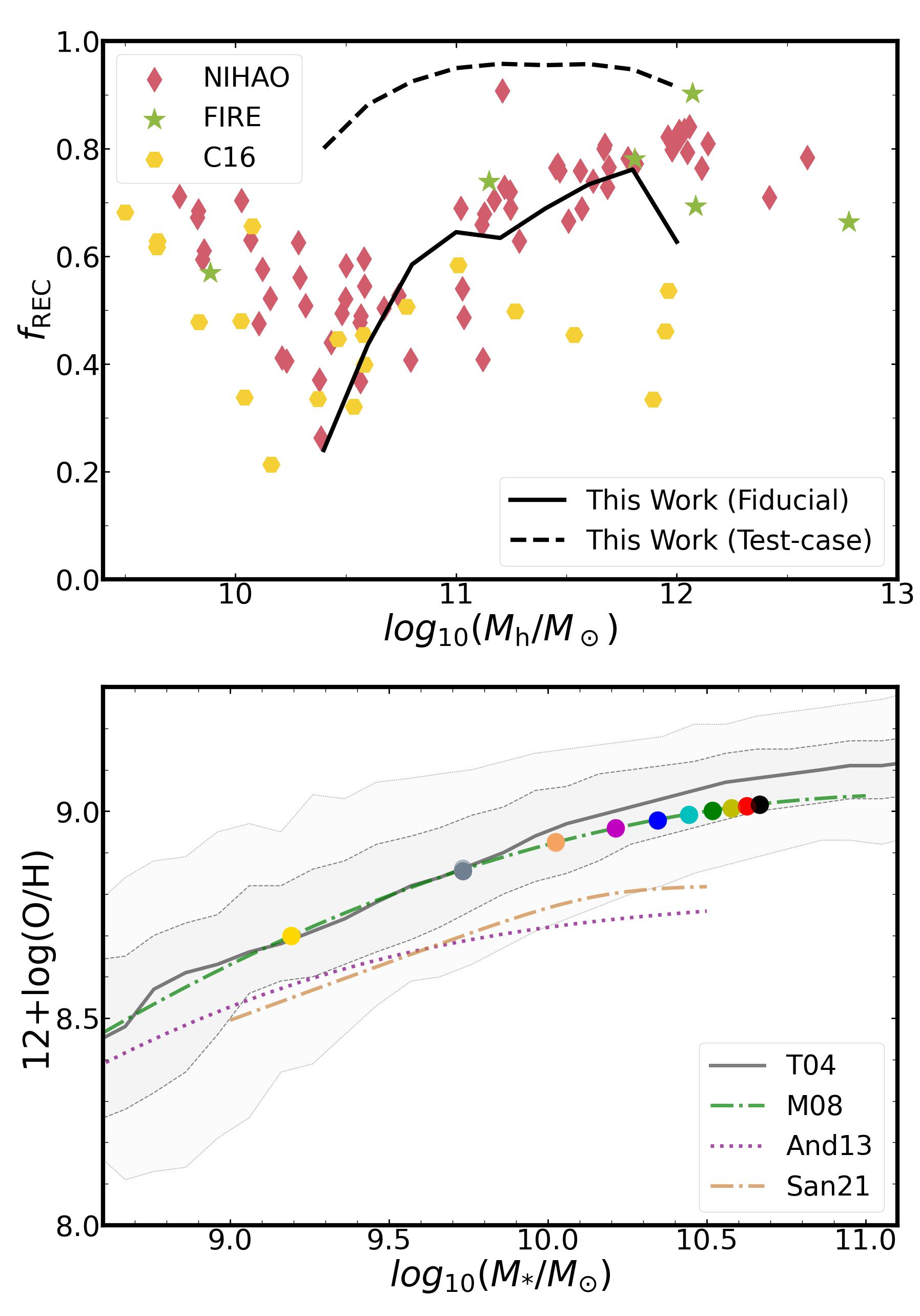}
    \caption{The best-fit results for parameter $f_\mathrm{REC}(M_\mathrm{h})$ and the corresponding predicted gas-phase mass-metallicity relation (MZR) at $z=0$.
    Upper panel: the best fitting results for parameter $f_\mathrm{REC}(M_\mathrm{h})$ for two different mass loading factors (black lines). 
    The solid line represents the case with the fiducial mass loading factor, while the dashed line corresponds to the test mass loading factor. 
    The data from hydrodynamical simulations is also plotted for comparison, including NIHAO simulation \citep{2019MNRAS.485.2511T}, FIRE simulation \citep{2017MNRAS.470.4698A} and C16 simulation \citep{2016ApJ...824...57C}. Lower panel: the gas-phase MZR at $z=0$. The model predictions under two different mass loading factors are shown as the color circles with different transparencies. 
    %These two models exhibit a high level of consistency with the observed MZR from \citet{2008A&A...488..463M}, resulting in a significant overlap between these two predictions. 
    All lines and shaded region are from different observational data. The solid, dashed, and dotted gray lines show the median, $68\%$ contour, and $95\%$ contour, respectively, of the MZR at $z\sim 0.1$ from \citet{2004ApJ...613..898T}. The green dash-dotted line corresponds to the MZR at $z=0.07$ from \citet{2008A&A...488..463M}, the purple dotted line represents the MZR at $z=0.027\sim 0.25$ from \citet{2013ApJ...765..140A}, and the brown dash-dotted line shows the MZR at $z\sim 0$ from \citet{2021ApJ...914...19S}.
    }
    \label{fig:Recycle_fraction}
\end{figure}

We utilize the observed gas-phase mass-metallicity relation (MZR) at $z\sim 0$ to constrain the recycle fraction $f_\mathrm{REC}$, based on the evolution of metals in ISM as described by Equation \ref{eq:metal_final}. 
Several well-established observational results of the local ($z\sim 0$) mass-metallicity relation (MZR) are available, as shown in the lower panel of Figure \ref{fig:Recycle_fraction}. For our calibration, we select the MZR from \cite{2008A&A...488..463M} as our benchmark, given its close agreement with measurements by \cite{2004ApJ...613..898T}. This choice implies that our model parameters are specifically optimized to reproduce the \cite{2008A&A...488..463M} data at $z\sim 0$.
The MCMC method is used to obtain the best-fit results for $f_\mathrm{REC}$, and we also explore the model with two mass loading factors, respectively: fiducial mass loading factor and test case (Full details are provided in Section \ref{subsec:method_constrain_baryon}).

The upper panel of Figure \ref{fig:Recycle_fraction} shows the best-fit results for parameter $f_\mathrm{REC}(M_\mathrm{h})$ for two different mass loading factors, with the dashed/solid black lines for the higher/lower mass loading factors, respectively. We also plot results from a few hydrodynamical simulations using different symbols in the figure.
Obviously, when applying the fiducial mass loading factor $\eta_\mathrm{m}$ (solid line), the obtained recycle fraction $f_\mathrm{REC}$ is $\sim25\%-75\%$ within the halo mass range of $10^{10.4}-10^{12.0}M_\odot$. This relation between $f_\mathrm{REC}$ and $M_\mathrm{h}$ aligns well with the results of hydrodynamical simulations, including the NIHAO simulation \citep{2019MNRAS.485.2511T}, the FIRE simulation \citep{2017MNRAS.470.4698A}, and the C16 simulation \citep{2016ApJ...824...57C}.
Furthermore, in Figure \ref{fig:Mass_loading_factor}, it is seen that the fiducial mass loading factor ($\eta_\mathrm{m}$ from the FIRE-2 simulation) is more consistent with observations. Therefore, it is reasonable to conclude that in our model, the mass loading factor derived from the FIRE-2 simulation \citep{2021MNRAS.508.2979P} is preferred. In addition, what is interesting is that no matter which mass loading factor is used in our work, the trend of recycle fraction with halo mass is similar. Specifically, $f_\mathrm{REC}$ first increases and then flattens or even decreases as the halo mass exceeds $10^{12}M_{\rm \odot}$.  It is also interesting to find that both the NIHAO \citep{2019MNRAS.485.2511T} and FIRE \citep{2017MNRAS.470.4698A} simulations indicate a similar trend, that is, $f_\mathrm{REC}$ has a peak around $M_\mathrm{h} \sim 10^{12}M_{\odot}$. 

Furthermore, the lower panel of Figure \ref{fig:Recycle_fraction} shows that no matter which mass loading factor is used in the model, the gas-phase MZR at $z=0$ predicted by the best-fit parameters in $f_\mathrm{REC}$ aligns perfectly with the observed results of \cite{2008A&A...488..463M}, with more discussion on mass-metallicity relation in Section \ref{subsec:result_prediction} Overall, Figure \ref{fig:Recycle_fraction} shows the effectiveness of gas-phase MZR in constraining the recycling process, at least for the MZR from \cite{2008A&A...488..463M}.

\subsection{Accretion fraction \texorpdfstring{$\epsilon$}{epsilon}} \label{subsec:result_best_fitting_acc}

\begin{figure}
    \centering
    \includegraphics[width=\columnwidth]{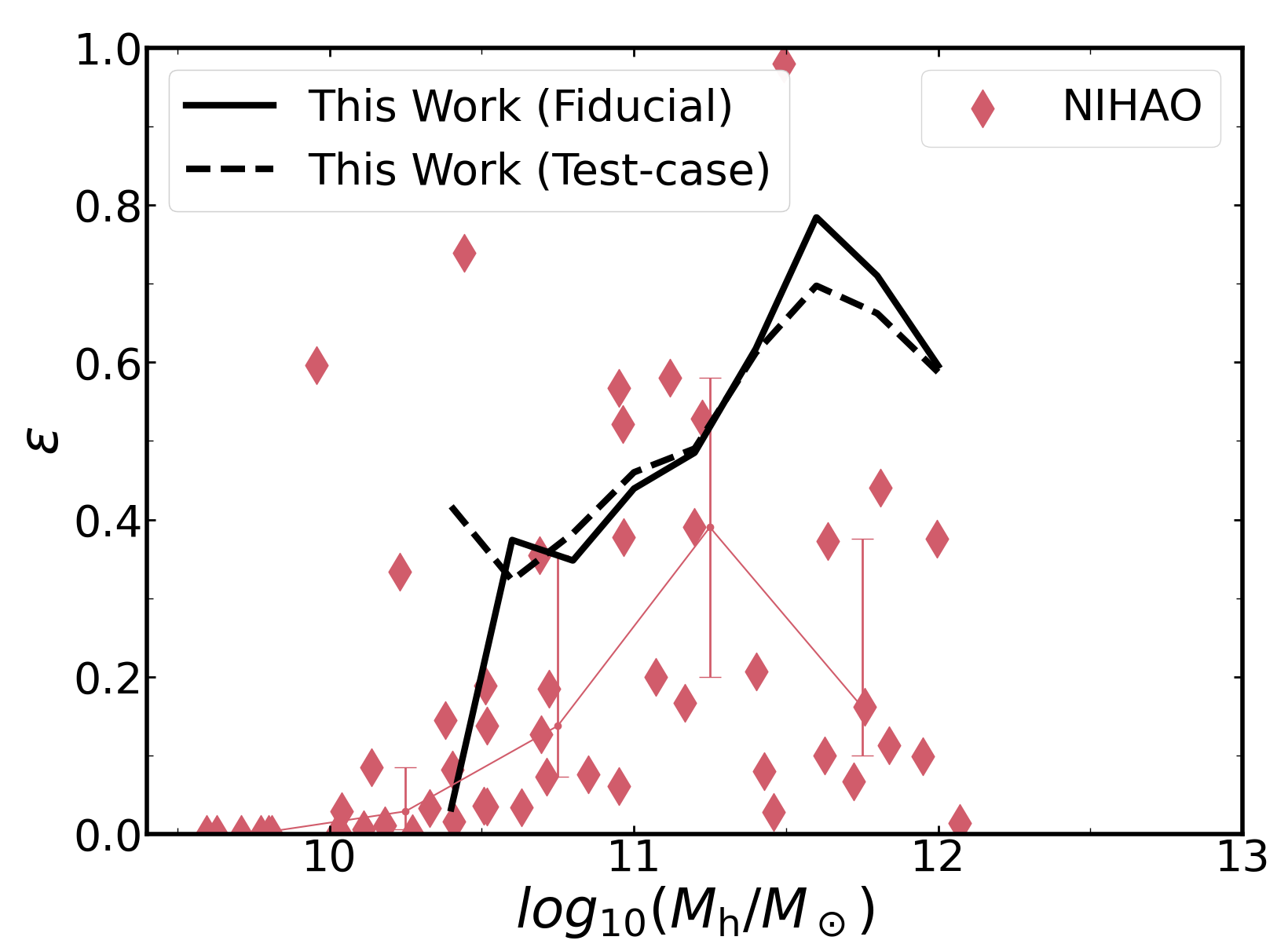}
    \caption{
    The best-fit results for accretion fraction $\epsilon(M_\mathrm{h})$ in models under two different mass loading factors (black lines). The red diamonds are from the NIHAO simulation, with red line and error-bars showing the medium and the scatter for NIHAO data.
    }
    \label{fig:Accretion_fraction}
\end{figure}

After obtaining the best-fit result for recycle fraction $f_\mathrm{REC}$, Equation \ref{eq:total_acc_rec} is then used to get the gas accretion rate from the IGM, $\dot{M}_\mathrm{gas,acc}$, so the parameter $\epsilon(M_\mathrm{h})$ in Equation \ref{eq:acc} can be obtained. Figure \ref{fig:Accretion_fraction} presents the best-fit results for parameter $\epsilon(M_\mathrm{h})$, and again the results from two different mass loading factors are shown as black lines. Here we remind the readers that $\epsilon(M_\mathrm{h})$ describes the instantaneous rate of gas accreted to the galaxy divided by the universal gas accretion from the IGM due to halo growth. Unfortunately, this quantity has not been well studied in the literature. Some studies \citep{2016ApJ...824...57C,2019MNRAS.485.2511T} have obtained the accumulative accretion fraction of gas over the life time of their model galaxies. Here we analyze the particle data from the NIHAO simulation and calculate the fresh accretion fraction from $z=0.1$ to $z=0$, which are represented by the red diamonds in Figure \ref{fig:Accretion_fraction}.

It is evident from the plot that regardless of which of the two loading factors we use, the best-fit parameter $\epsilon(M_\mathrm{h})$ is very similar. This is understandable because for a higher mass loading factor, a higher recycle fraction $f_\mathrm{REC}$ is required by the gas-phase MZR (shown in the previous section) to balance the gas content. The black lines show that the accretion fraction of gas from the IGM is decreasing with halo mass, and it is around $40\%$ at the halo mass about $10^{11.5}M_{\rm \odot}$. As mentioned before, the suppress of gas accretion from IGM is mainly caused by two effects. One is the cosmic re-ionization and the other is the preventive effect due to feedback gas escaping from the galaxy into the IGM. However, for the re-ionization effect, it is negligible in galaxy with halo mass greater than $10^{10}M_{\rm \odot}$. Our results show that it is mainly the preventive effect that suppresses the gas accretion into halo from the IGM. 

The red diamonds in Figure \ref{fig:Accretion_fraction} show that the gas accretion fraction in the NIHAO simulation has a peak at $M_\mathrm{h}\sim 10^{11}M_{\rm \odot}$ although the scatter is rather large. Compared to the NIHAO simulation, our model prediction roughly agrees with the simulation results, while the prediction is higher than that of the simulation for massive halos, showing that the preventive effect in NIHAO simulation is stronger for massive galaxies. However, one possible explanation could also be the ineffective cooling of the hot gas in massive galaxies above $10^{12}M_{\rm \odot}$, where the cooling regime is dominant by the hot mode.

\begin{figure}
    \centering
    \includegraphics[width=\columnwidth]{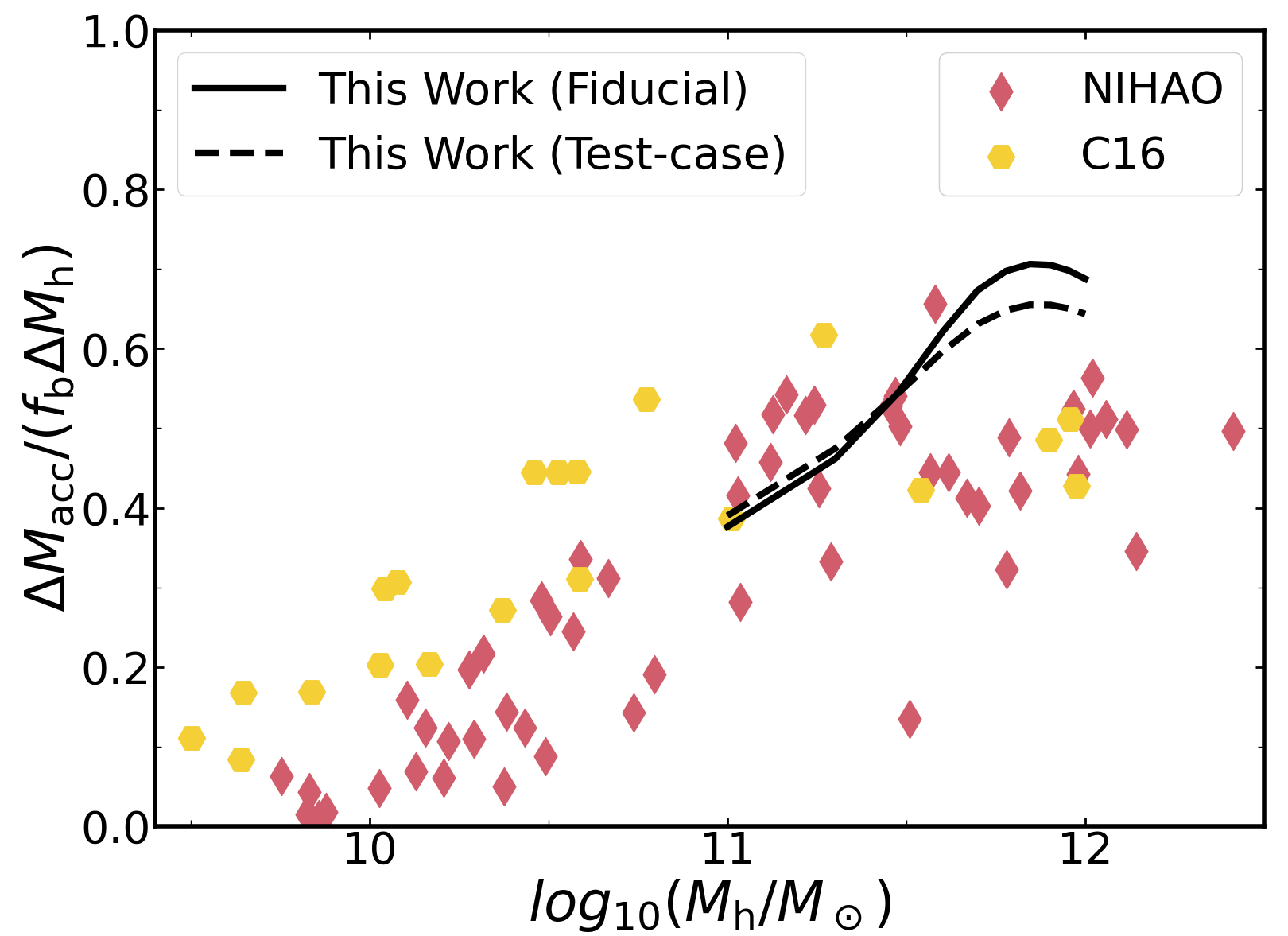}
    \caption{The accumulative accretion fraction for our model galaxies, from $z=2$ to $z=0$. The results of our model are shown as the solid black line (fiducial mass loading factor) and dashed black line (test case), respectively. The simulation results are shown as red diamonds (NIHAO simulation, \citet{2019MNRAS.485.2511T}) and yellow hexagons (C16, \citet{2016ApJ...824...57C}).
    }
    \label{fig:Cumulative_accretion}
\end{figure}

We also calculate the accumulative accretion fraction from $z=2$ to $z=0$, $\Delta M_\mathrm{acc}/(f_\mathrm{b}\Delta M_\mathrm{h})$, for our model galaxies, enabling a more direct comparison with results from simulations. The results are presented in Figure \ref{fig:Cumulative_accretion}.
We find that for our model galaxies, the general trend of accumulate accretion fraction with the halo mass at $z=0$ agree with results from a few simulations. Specifically, the accumulative accretion fraction gradually increases as the halo mass at $z=0$ increases from $10^{11}M_\odot$ to $\sim 10^{11.5}-10^{11.6}M_\odot$. The fraction flattens and subsequently decreases when the halo mass exceeds $\sim 10^{11.5}-10^{11.6}M_\odot$.
However, similar to the accretion fraction $\epsilon$ shown in Figure \ref{fig:Accretion_fraction}, our model predicts a higher accumulative accretion fraction at the high-mass end compared to these simulations.

\subsection{Predictions on metal content and evolution} \label{subsec:result_prediction}

In previous sections, we have determined the best-fit parameters for the recycle fraction ($f_\mathrm{REC}$) and the accretion fraction ($\epsilon$). More importantly, combining Figure \ref{fig:Mass_loading_factor} and Figure \ref{fig:Recycle_fraction}, it is natural to conclude that the fiducial mass loading factor from the FIRE-2 simulation \citep{2021MNRAS.508.2979P} is more preferred. 
We now employ our fiducial model to generate predictions about the metal content in the galaxy and the evolution with redshift.

It is noteworthy that all metal-related relations can be predicted solely using the recycle fraction, $f_\mathrm{REC}$. This includes the gas-phase MZR at various redshifts (Section \ref{subsubsec:result_prediction_gasMZR_at_gighz}), the stellar mass-metallicity relation (Section \ref{subsubsec:result_prediction_stellarMZR}), the metallicity in the CGM (Section \ref{subsubsec:result_prediction_mass_metallicity_CGM}), and the distribution of metals across different components (Section \ref{subsubsec:result_prediction_metal_budget}).
In our model, once the evolution of stellar and cold gas content in a model galaxy is determined, the metal evolution is entirely governed by the combination of the outflow and recycling processes. In other words, the metal evolution depends on the mass loading factor ($\eta_\mathrm{m}$), the recycle fraction ($f_\mathrm{REC}$), and the recycle time ($t_\mathrm{REC}$), as discussed frequently in previous sections.

\subsubsection{The gas-phase MZR at high-z}
\label{subsubsec:result_prediction_gasMZR_at_gighz}

\begin{figure}
    \centering
    \includegraphics[width=\columnwidth]{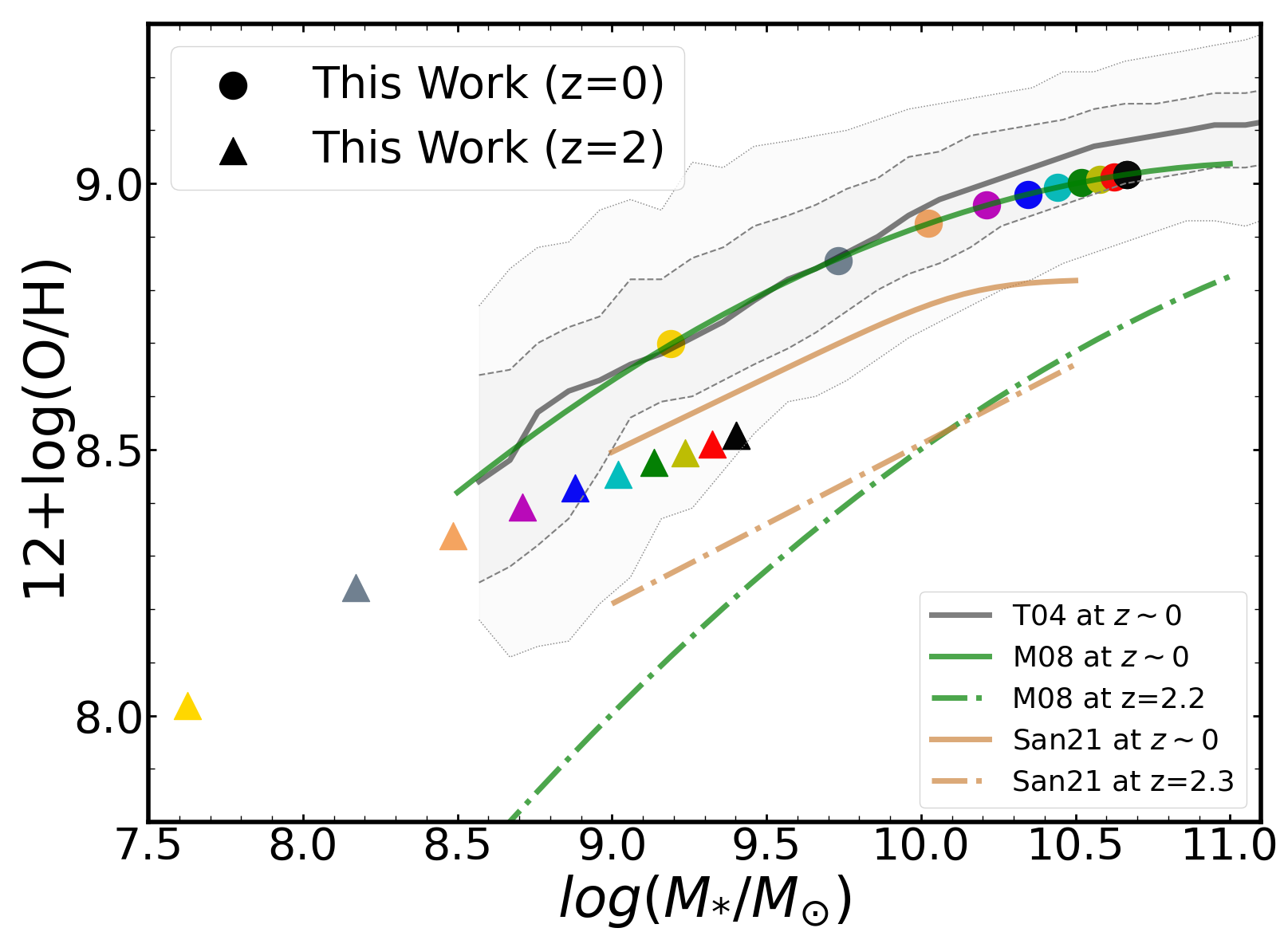}
    \caption{
    The gas-phase mass-metallicity relation (MZR) at $z=2$. Model prediction: The MZRs predicted by our fiducial model are shown as colored scatter points. Different shapes represent different redshifts, such as circles for $z=0$ and triangles for $z=2$.
    Observations: The observed results for gas-phase MZR are shown as colored lines, as labeled in the figure. 
    In brief, the solid lines represent the MZRs at $z\sim 0$ from observation, including \citet{2004ApJ...613..898T}, \citet{2008A&A...488..463M} and \citet{2021ApJ...914...19S}. The green dashed-dotted line represents the MZR from \citet{2008A&A...488..463M} at $z=2.2$ and the brown dashed-dotted line represents the MZR from \citet{2021ApJ...914...19S} at $z=2.3$.
    }
    \label{fig:Gas_MZR_at_high_redshift}
\end{figure}

In this part, we examine the performance of the fiducial model in predicting the gas-phase MZR at $z=2$ and compare our prediction with the observed MZRs from different observations \citep{2008A&A...488..463M,2021ApJ...914...19S}. One big issue with the measured MZRs is that there is obvious discrepancy among different observations. Taking the MZRs at $z=0$ shown in the lower panel of Figure \ref{fig:Recycle_fraction} as an example, the results of \cite{2004ApJ...613..898T} and \cite{2008A&A...488..463M} are relatively close, but differ significantly from those of \cite{2021ApJ...914...19S} and \cite{2013ApJ...765..140A}. This discrepancy arises from the different sets of metallicity calibrations used to derive the MZR, as detailed in Section 6.1.2 of \cite{2021ApJ...914...19S}. A comprehensive discussion of these differences is beyond the scope of this study.

In Figure \ref{fig:Gas_MZR_at_high_redshift}, the colored points represent our prediction for the gas-phase MZR at $z=0$ (circles) and $z=2$ (triangles). These predictions are compared with observational data from \cite{2008A&A...488..463M} and \cite{2021ApJ...914...19S}. Our MZR prediction at $z=0$ shows excellent agreement with the local universe observations of \cite{2008A&A...488..463M}, as expected since these data were used to calibrate our model parameters. At a higher redshift of $z=2$, our model predicts systematically higher gas-phase metallicities compared to the MZR reported by \cite{2008A&A...488..463M} at $z=2.2$ and \cite{2021ApJ...914...19S} at $z=2.3$. The tension between our results and observations at $z\sim 2$ could originate from: (i) incomplete physical treatment in the current model, and (ii) potential systematics in metallicity measurements. We will conduct a detailed analysis and corresponding model refinements to address this in future work.

\subsubsection{The stellar mass-metallicity relation} \label{subsubsec:result_prediction_stellarMZR}

It has been well demonstrated that there is a relatively tight relation between stellar metallicity and stellar mass of galaxies \citep[e.g.][]{2005MNRAS.362...41G,2013ApJ...779..102K,2024arXiv240112310G}, which can also be used to constrain the model of galaxy formation. As shown in previous sections, the observed gas-phase mass-metallicity relation ($\mathrm{MZR_{gas}}$) at $z=0$ is used to constrain the model parameters, here we use the stellar mass-metallicity relation ($\mathrm{MZR_{*}}$) as an independent test of our model prediction. In our model, the metal content in star can be calculated using the following equation:
\begin{equation}
    \dot{M}_\mathrm{Z,*}(t)=Z_\mathrm{gas}(t)\times \dot{M}_*(t).
	\label{eq:stellar_metal}
\end{equation}
Then with the definition of stellar metallicity ($Z_*=M_\mathrm{Z,*}/M_*$), we can easily derive the evolution of stellar metallicity and naturally obtain the $\mathrm{MZR_{*}}$ at $z=0$.

\begin{figure}
    \centering
    \includegraphics[width=\columnwidth]{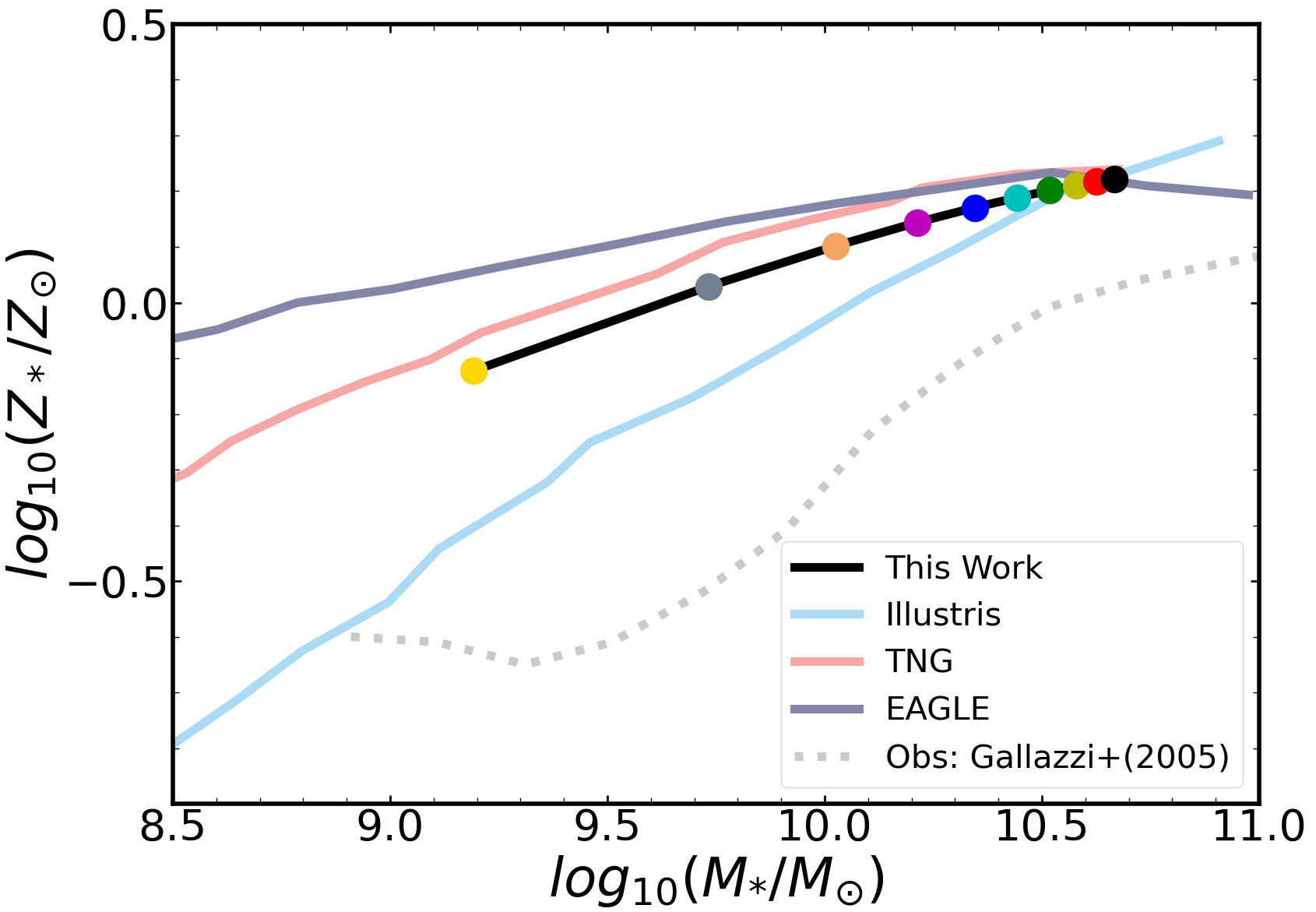}
    \caption{
    The stellar mass-metallicity relation ($\mathrm{MZR_*}$) at $z=0$. The prediction of our fiducial model for $\mathrm{MZR_*}$ at $z=0$ is shown as color circles connected by the solid black line. The light color lines correspond to results from \citet{2024arXiv240112310G} for Illustris simulation (blue), TNG simulation (pink), and EAGLE simulation (purple).The dashed gray line represents the observation of $\mathrm{MZR_*}$ at $z=0$ from \citet{2005MNRAS.362...41G}.
    }
    \label{fig:Stellar_MZR}
\end{figure}

The results of $\mathrm{MZR_{*}}$ at $z=0$ for our fiducial model are shown as color circles connected by a solid black line in Figure \ref{fig:Stellar_MZR}. We also show the predictions obtained by \cite{2024arXiv240112310G} from a few state-of-the-art hydrodynamical simulations. As can be seen, $\mathrm{MZR_{*}}$ in our model exhibits a trend of increasing metallicity with increasing stellar mass, but the relation becomes flat at high mass end. This behavior is consistent with previous observations \citep[e.g.][]{2005MNRAS.362...41G,2013ApJ...779..102K}. In addition, the normalization of the $\mathrm{MZR_{*}}$ in our work falls within the range of predictions for the Illustris, TNG and EAGLE simulations. 
Clearly, all these theoretical predictions are higher than the observations of \cite{2005MNRAS.362...41G}. As discussed in Section 3.1 of \cite{2024arXiv240112310G}, this discrepancy can be attributed to the different computing methods employed in the theoretical model and the observations. Finally, we also test the predictions of our model with different mass loading factors and find no significant differences in the predicted $\mathrm{MZR_{*}}$ among these models, albeit the results are not shown here.

\subsubsection{The mass and metallicity of the circumgalactic medium }\label{subsubsec:result_prediction_mass_metallicity_CGM}

\begin{figure*}
    \centering
	\includegraphics[scale=0.23]{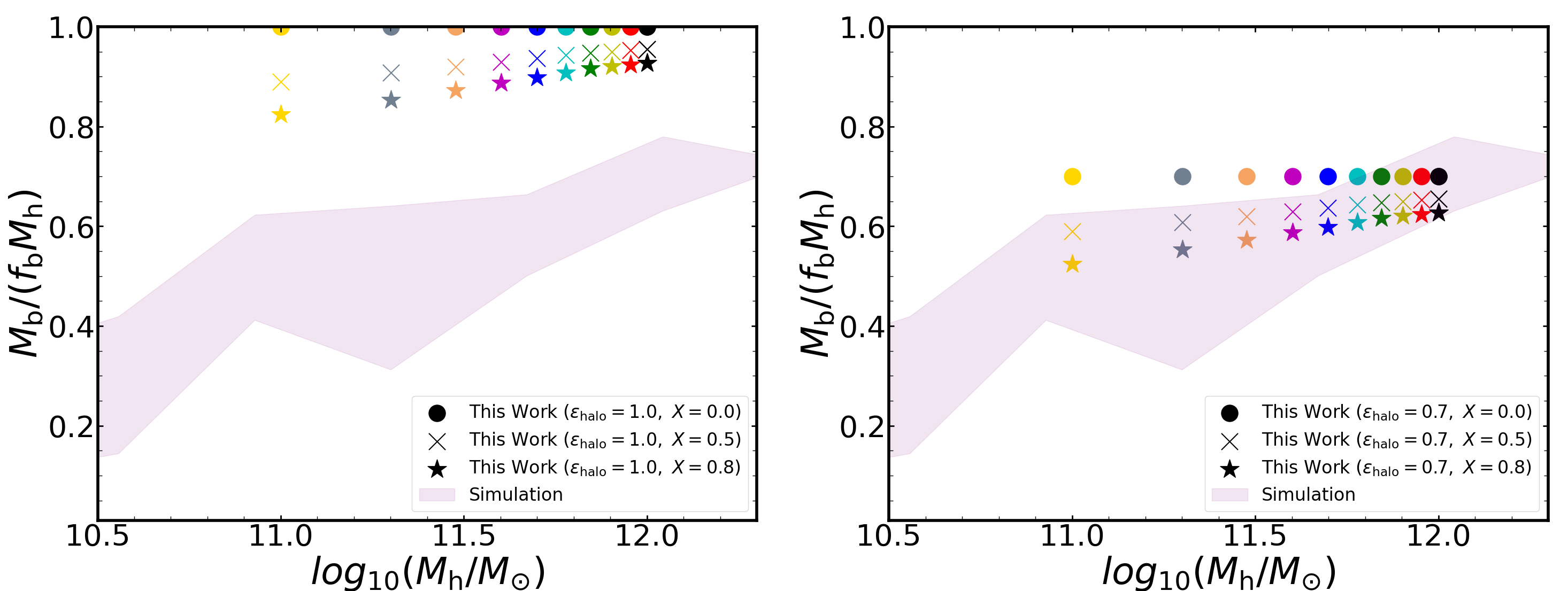}
    \caption{The total baryon fraction within halo (scaled by the cosmic baryon fraction) as a function of halo mass at $z=0$. The prediction of our model are presented by colored scatters in both panels, but with different parameters (detailed in text). The result of simulations is shown as the light purple shading.}
    \label{fig:Baryon_fraction_in_halo}
\end{figure*}

\begin{figure}
    \centering
    \includegraphics[width=\columnwidth]{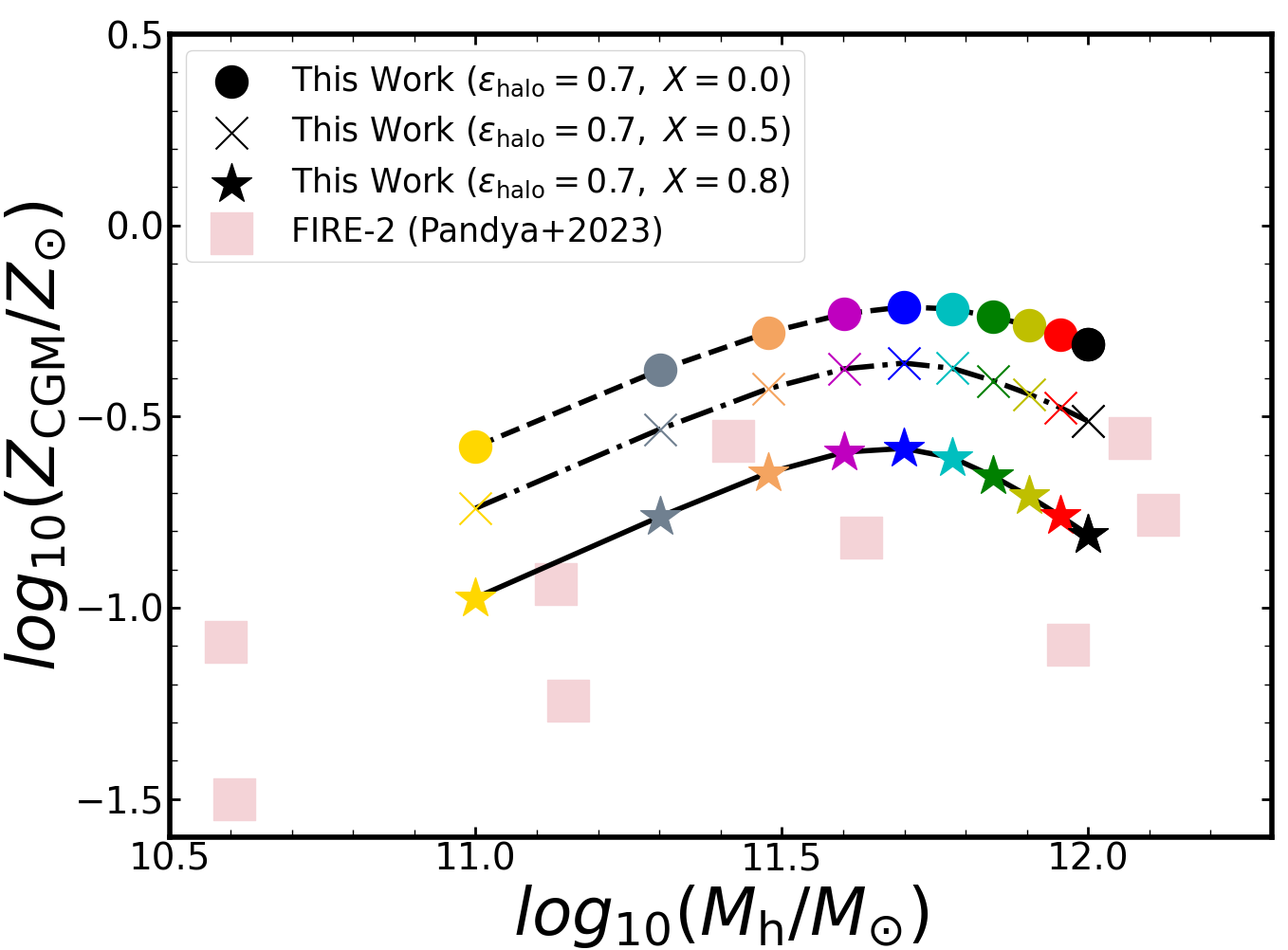}
    \caption{
    The metallicity in the circumgalactic medium (CGM) as a function of halo mass at $z=0$. The predictions of our model are represented by colored circles ($\epsilon_\mathrm{halo}=0.7,\ X=0.0$), crosses ($\epsilon_\mathrm{halo}=0.7,\ X=0.5$) and stars ($\epsilon_\mathrm{halo}=0.7,\ X=0.8$) connected by various lines, respectively. The result from \citet{2023ApJ...956..118P} is shown as pink squares for the FIRE-2 simulation. 
    }
    \label{fig:CGM_metallicity}
\end{figure}

The circumgalactic medium (CGM) acts as a critical interface between the galaxy and its local environment, playing a key role in regulating star formation and the overall baryon cycle . 
Specifically, the CGM serves as a gas reservoir for star formation in the galaxy, a venue for galactic feedback and recycling, and potentially a key regulator of the galaxy's gas supply \citep[e.g.][]{2017ARA&A..55..389T,2023ARA&A..61..131F}. 
Therefore, by studying the properties of CGM (e.g. total gas content and metallicity), one can gain insights into the history of gas accretion and ejection around galaxies, and the efficiency of different feedback mechanisms \citep[e.g.][]{2017MNRAS.468.4170M,2018MNRAS.473..538C,2019MNRAS.488.1248H,2023MNRAS.525.5677S,2024ApJ...977..233C,2024arXiv240700172R}. 
Unfortunately, measuring the properties of CGM through observations presents significant challenges due to its diffuse and low-density nature. 
Current measurements of gas content in the CGM are primarily based on X-ray emission \citep[e.g.][]{2020ApJ...897...63D,2023ApJ...955L..21N,2024A&A...690A.267Z} and the thermal SZ effect \citep[e.g.][]{2022ApJ...928...14B,2023ApJ...951..125D}. However, these methods are typically applicable to massive galaxies (although \cite{2023ApJ...951..125D} does extend to a lower mass) and are influenced by modeling the CGM temperature and density distribution. 
For the mass and metal content of CGM in galaxies with a mass comparable to our sample ($10^{11}M_{\rm \odot}$ to $10^{12}M_{\rm \odot}$), most constraints are from hydrodynamical simulations \citep[e.g.][]{2016ApJ...824...57C,2019MNRAS.485.2511T,2019MNRAS.488.1248H,2023ApJ...956..118P}. In this section, we compare our model predictions for the mass and metallicity of CGM with results from simulations.

To predict the mass of the CGM, the first step is to determine the total baryon mass within the halo of a galaxy, which can be written as, 
\begin{equation}
    M_\mathrm{b}=\epsilon_\mathrm{halo}f_\mathrm{b}M_\mathrm{h}- X\int(1-f_\mathrm{REC})\dot{M}_\mathrm{gas,out}dt.
	\label{eq:total_baryon_in_halo}
\end{equation}
The first term represents the total gas that has been accreted into the halo, characterized by an accretion parameter $\epsilon_\mathrm{halo}$. We again remind the readers that $\epsilon(M_\mathrm{h})$ in Equation \ref{eq:acc} parameterizes the final gas accretion to the galaxy, while here $\epsilon_\mathrm{halo}$ describes the gas accreted to the halo. Unfortunately, the current constraint on $\epsilon_\mathrm{halo}$ is also weak, while a few simulations \citep[e.g.][]{2016ApJ...824...57C,2019MNRAS.485.2511T} have found that $\epsilon_\mathrm{halo}$ is approximately $0.9\sim 1$ for a galaxy with halo mass between $10^{11}M_\odot\sim 10^{12}M_\odot$, and it gradually decreases to $0.5\sim 0.7$ as the halo mass drops to $\sim 3\times 10^{10}M_\odot$. The term $\int(1-f_\mathrm{REC})\dot{M}_\mathrm{gas,out}dt$ gives the total gas in the outflow that is never recycled, and $X$ specifies how much of this gas is ejected out of the halo, with $X=1.0$ means that $100\%$ of this gas is ejected out of the halo, and $X=0.0$ means that all the gas remains in the CGM.

In this study, we test a few cases for $\epsilon_\mathrm{halo}$ and $X$ with $\epsilon_\mathrm{halo}=1.0$, $\epsilon_\mathrm{halo}=0.7$ and $X=0.0,\ 0.5,\ 0.8$. The results of the total baryon fraction are shown in Figure \ref{fig:Baryon_fraction_in_halo}. For comparison, the purple shading area shows the simulation results at $z=0$ from \cite{2016ApJ...824...57C}, \cite{2019MNRAS.485.2511T}, as well as that from \cite{2019MNRAS.488.1248H} at $z=0.25$. The total baryon fraction in these simulations exhibits a clear dependence on the halo mass. However, our model predicts a weaker dependence on halo mass. This is simply because we use a constant $\epsilon_\mathrm{halo}$, while simulations do find some mass dependence. The left panel clearly shows that the predicted baryon fraction with $\epsilon_\mathrm{halo}=1.0$ is significantly higher than that from simulations, regardless of which value of $X$ is adopted. It is found from the right panel that a model with $\epsilon_\mathrm{halo}=0.7$ and $X=0.8$ agrees better with the simulation results.

Once the total baryon mass and the fraction of outflow that escapes from the halo are determined, the total mass in CGM can be obtained as $M_\mathrm{CGM}=M_\mathrm{b}-M_*-M_\mathrm{gas}$, where $M_\ast$ is the stellar mass, and $M_\mathrm{gas}$ is the cold gas mass within the galaxy. In terms of the metal mass, we assume that the metal flow has the same fate as the gas in the outflow. This allows us to derive the metallicity of CGM. In Figure \ref{fig:CGM_metallicity} we show the metallicity of the CGM as a function of halo mass, and we also present the prediction from the FIRE-2 simulation \citep{2023ApJ...956..118P} (pink squares). It is seen that both our model and the FIRE-2 simulation predict the same trend, that the metallicity of CGM initially increases with halo mass, and it becomes flat at the massive end. The peak metallicity occurs at a halo mass of approximately $\sim 10^{11.5} M_\odot-10^{11.7} M_\odot$.
Here we show predictions with three sets of parameters: $\epsilon_\mathrm{halo}=0.7$ with $X=0.0$, $X=0.5$, and $X=0.8$, respectively. It is shown that the scenario with $X=0.0$ (all the never-recycled metals and gas stay in the CGM) naturally provides an upper limit of the metallicity of CGM. As $X$ increases, the predicted metallicity gradually decreases, approaching the result of the FIRE-2 simulation. It shows the parameter set of $\epsilon_\mathrm{halo}=0.7$ and $X=0.8$ best fits the FIRE-2 simulation data. Future robust observational data on CGM metallicity are expected to provide further constraints on our model. We note that the above analysis is based on the assumption that the metal in outflow entirely follows the gas in outflow. If the gas in the outflow is metal richer than the ISM or the metal-enriched outflow has a different fate, the constraints on parameter $X$ will be different, but discussion on this issue is beyond the scope of our current work.

\subsubsection{Metal budget}\label{subsubsec:result_prediction_metal_budget}

\begin{figure}
    \centering
    \includegraphics[width=\columnwidth]{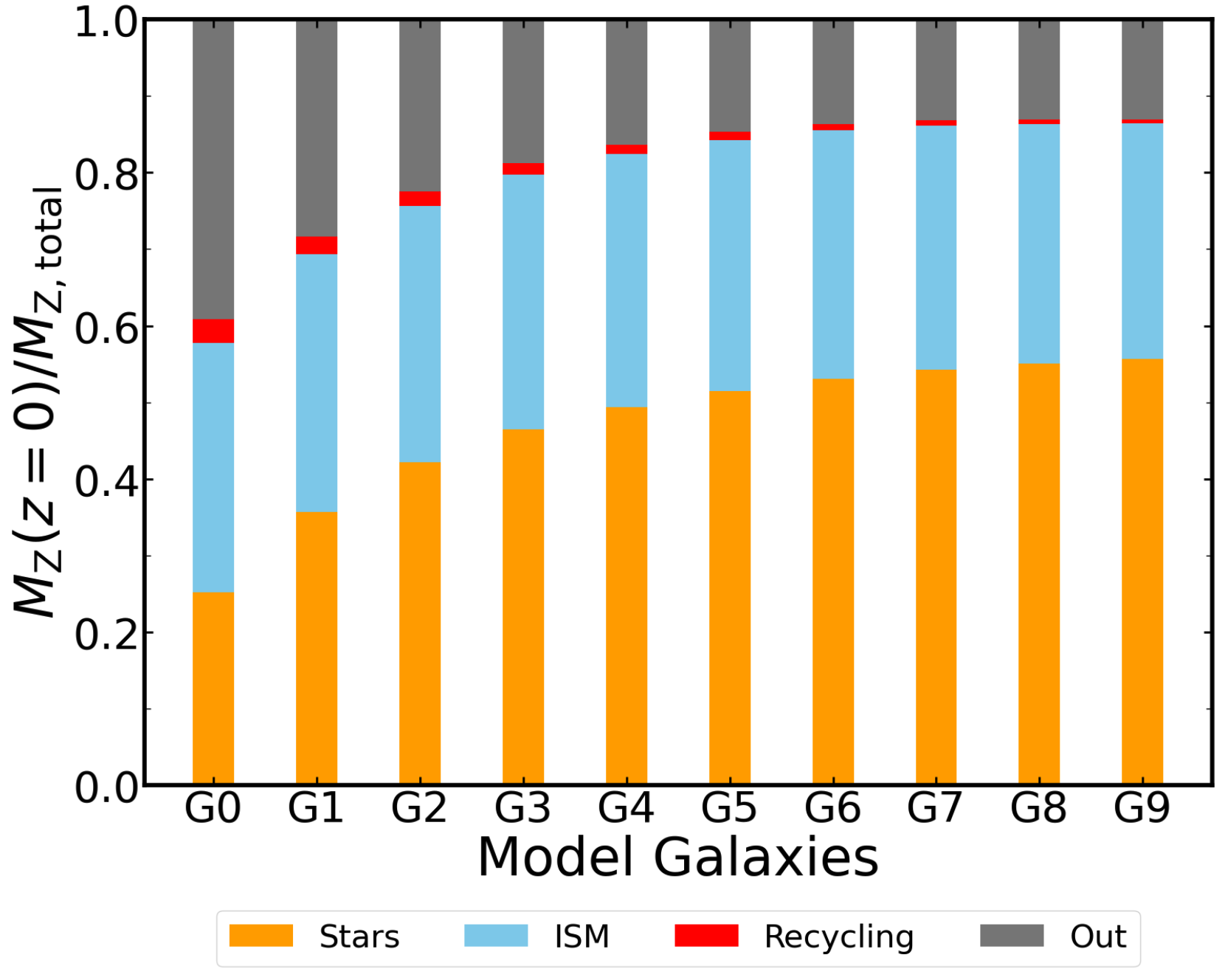}
    \caption{Amount of metals in each component at $z=0$ for model galaxies in fiducial model, including the metal locked in stars (orange), in ISM (blue), undergoing recycle process (red) and that will never re-accreted into the galaxy again (gray).}
    \label{fig:Metal_budget}
\end{figure}

It has been shown that the metal budget, its fraction in different baryonic components, can be used to constrain the model for galaxy formation \citep[e.g.][]{2017MNRAS.468.4170M,2022MNRAS.511.2600M}. 
Our model for baryon cycle specifies how gas and metal are cycled inside the galaxy. Then it is interesting to investigate the metal budget. Here we examine the final distribution of metals in our model galaxies at $z=0$, and we categorize the metal content into four components: those locked by the stars ($Stars$); those in the ISM ($ISM$); those that have been blown out and are undergoing the recycle process ($Recycling$); those that have been blown out and will never return/recycled to the galaxy/ISM ($Out$). Note that the $Out$ component in this work refers to the metals that have been expelled from the ISM, reach the CGM/IGM and stay in the CGM/IGM permanently. In short, this part of the metal will not return/recycle to the ISM to participate in further star formation.
Figure \ref{fig:Metal_budget} shows the proportion of metals in each component at $z=0$ to the total metal produced by galaxies, that is, $\frac{M_\mathrm{Z}(z=0)}{M_\mathrm{Z,total}}$. 

Obviously, for our model galaxy, the proportion of metal locked in stars increases with the mass of the galaxy. 
For galaxy $G9$, which has a halo mass of $M_\mathrm{h}(z=0)=10^{12}M_\odot$, the metal fraction in the stellar component exceeds $50\%$. 
Similar results have been reported using hydrodynamical simulations. \cite{2017MNRAS.468.4170M} have found that in the FIRE simulation the portion of metals in stars is higher in massive galaxies, and the metal budgets of $\mathrm{L^*}$ galaxies at $z\sim 0$ are dominated by the stellar component, as high as $\sim 80\%$. It is also found in the EAGLE simulation that the metal fraction in the stellar component is $\sim 20\%$ for galaxies with a halo mass of $\sim 10^{11}M_\odot$, and this fraction rises to $\sim 40\%$ for galaxies with a halo mass of $\sim 10^{12}M_\odot$ \citep{2022MNRAS.511.2600M}.

Intriguingly, the metal fraction in cold gas or the ISM component is $\sim 30\%$ for all of the model galaxies, unlike the behavior observed in the stellar component. There is no consensus on the metal budget in the ISM. In galaxies analyzed in the FIRE simulation \citep{2017MNRAS.468.4170M}, the metal fraction in the ISM component varies randomly within the range of $10\%$ to $40\%$. 
And for galaxies with halo masses between $10^{11}M_\odot$ and $10^{12}M_\odot$ in the EAGLE simulation \citep{2022MNRAS.511.2600M}, the proportion of metals in the ISM component is roughly the same, which is a relatively low value of around $8\%-9\%$. 

Figure \ref{fig:Metal_budget} also shows that the proportion of metals in the $Recycling$ component is at a few percent level, and it becomes negligible in massive galaxies. This trend is probably attributed to the recycle time $t_\mathrm{REC}$ that decreases as the halo mass increases. As for the $Out$ component, the results of our work are similar to the studies based on the FIRE and EAGLE simulation \citep[e.g.][]{2017MNRAS.468.4170M,2022MNRAS.511.2600M}, showing that as galaxy mass increases, the proportion of metals in the $Out$ component decreases. This is expected because the gravitational potential of a low-mass halo is shallow, making it easier for gas and metals to be expelled by stellar feedback and winds.

\section{Summary} \label{sec:summary}

In this paper, we improve our previous empirical model \citep{2023MNRAS.519.1899C} to better constrain the baryon cycle in galaxies. We utilize the stellar mass-halo mass relations (SHMRs) at various redshifts and the halo growth history to obtain the star formation history (SFH) of model galaxies with halo masses ranging from $10^{11}M_\odot$ to $10^{12}M_\odot$ at $z=0$. We introduce the NeutralUniverseMachine model recently proposed by \citet{2023ApJ...955...57G} to determine the cold gas content and its evolution in the galaxy. Then we model the gas cycle processes in detail and use the constraints from both observations and simulations to refine the parameters in the models of the gas cycle. Our results can be summarized as the following, 

\begin{enumerate}

    \item The halo growth histories and the star formation history of model galaxies are almost the same as the results in \citet{2023MNRAS.519.1899C}. 

    \item Based on the NeutralUniverseMachine model, the cold gas content in our model galaxies is slightly increasing with the time from $z=2$ to $z=0$, but the gas fraction $\mu_\mathrm{gas}$ is decreasing over time, indicating that the increase of stellar mass slightly surpasses the cold gas. Using the NeutralUniverseMachine model, the obtained $\mu_\mathrm{gas}-M_*$ relations are generally consistent with the relations from the TNG50 simulation.
    
    \item Given the evolution of halo, star, and gas, along with the chemical evolution model of galaxies, and then selecting a mass loading factor, the observed gas-phase MZR at $z=0$ can effectively constrain the recycling process. Surprisingly, both the observations and our model favor the mass loading factor $\eta_\mathrm{m}$ derived from the FIRE-2 simulation \citep{2021MNRAS.508.2979P}.  
    With the model utilizing $\eta_\mathrm{m}$ from the FIRE-2 simulation as our fiducial model, we get a recycle fraction that first increases and then flattens or even decreases as the halo mass increases. This fraction varies approximately between $25\%$ and $75\%$, peaking around $M_\mathrm{h}\sim 10^{11.8}M_\odot$.

    \item The best-fit results for accretion fraction $\epsilon$ are similar in models with different mass loading factors. In general, the accretion fraction $\epsilon$ initially increases and then decreases with increasing halo mass, reaching the peak at around $10^{11.6}M_\odot$. When compared with other simulations, our results show a higher accretion fraction, particularly for massive galaxies.

    \item We present several predictions regarding the CGM and metal-related relations based on our fiducial model. Firstly, to fit the baryon mass with halo mass relation from hydrodynamical simulations, our model requires that the accretion of gas into the galaxy halo is around $70\%$ on average for halo with current mass between $10^{11}M_{\rm \odot}$ and $10^{12}M_{\rm \odot}$, and around $80\%$ of gas outflow has to escape the halo. Secondly, if metal has the same fate as the outflow gas, then the predicted metallicity of CGM also agrees with results from the FIRE-2 simulation. Finally, we predict the metal budget in four components: $Stars$, $ISM$, $Recycling$, and $Out$. The proportion of metals locked in $Stars$ increases with galaxy mass, reaching $55\%$ in galaxy with $M_\mathrm{h}(z=0)=10^{12}M_\odot$. In contrast, the metal fraction in the ISM remains around $\sim 30\%$ for all model galaxies.
    
\end{enumerate}

The main goal of this work is not to seek a complete and comprehensive model for gas cycle process, but to provide physical insight into galaxy formation and evolution, particularly focusing on the baryon cycle process. Fortunately, our model has successfully established a direct link between the gas recycling process and the evolution of metal, and we hope our parameterization of related physical processes, such as gas recycle fraction and the fraction of outflow escaping the halo, can be applied in semi-analytic model to study galaxy formation and evolution.

\begin{acknowledgments}
This work is partly supported by the National Key Research and Development Program (No. 2022YFA1602903, No. 2023YFB3002502, No. 2020SKA0110100), the science research grants from the China Manned Space project with NO.CMS-CSST-2025-A10. Xi Kang acknowledges the hospitality of the International Center of Supernovae (ICESUN), Yunnan Key Laboratory at Yunnan Observatories Chinese Academy of Sciences. HG acknowledges the support of the CAS Project for Young Scientists in Basic Research (No. YSBR-092),  and GHfund C(202407031909). Hou-Zun Chen acknowledges the cosmology simulation database (CSD) in the National Basic Science Data Center (NBSDC) and its funds the NBSDC-DB-10. We also acknowledge the support of the National Natural Science Foundation of China (No. 12403016, 12233005) and the Postdoctoral Fellowship Program of CPSF (No. GZC20241514).
\end{acknowledgments}

\bibliography{reference}{}
\bibliographystyle{aasjournal}

\end{CJK*}
\end{document}